\documentclass[12pt]{emulateapj}

\shorttitle{Dynamics of Merging Clusters}
\shortauthors{Dawson}

\usepackage[]{natbib}
\usepackage{amsmath}

\begin{document}


\title{The Dynamics of Merging Clusters:\\
		A Monte Carlo Solution Applied to the Bullet and Musket Ball Clusters}


\author{William A. Dawson}
\affil{University of California, Davis,Physics Department, One Shields Av., Davis, CA 95616, USA}
    
\email{wadawson@ucdavis.edu}



\begin{abstract}

Merging galaxy clusters have become one of the most important probes of dark matter, providing evidence for dark matter over modified gravity and even constraints on the dark matter self-interaction cross-section.  
To properly constrain the dark matter cross-section it is necessary to understand the dynamics of the merger, as the inferred cross-section is a function of both the velocity of the collision and the observed time since collision.
While the best understanding of merging system dynamics comes from N-body simulations, these are computationally intensive and often explore only a limited volume of the merger phase space allowed by observed parameter uncertainty.  
Simple analytic models exist but the assumptions of these methods invalidate their results near the collision time, plus error propagation of the highly correlated merger parameters is unfeasible.
To address these weaknesses I develop a Monte Carlo method to discern the properties of \emph{dissociative mergers} and propagate the uncertainty of the measured cluster parameters in an accurate and Bayesian manner.
I introduce this method, verify it against an existing hydrodynamic N-body simulation, and apply it to two known dissociative mergers: \object{1ES 0657-558} (\object{Bullet Cluster}) and \object{DLSCL J0916.2+2951} (\object{Musket Ball Cluster}).
I find that this method surpasses existing analytic models --- providing accurate (10\% level) dynamic parameter and uncertainty estimates throughout the merger history.
This coupled with minimal required \emph{a priori} information (subcluster mass, redshift, and projected separation) and relatively fast computation ($\sim$6 CPU\,hours) makes this method ideal for large samples of dissociative merging clusters.

\end{abstract}


\keywords{galaxies: clusters: individual (1ES 0657-558), galaxies: clusters: individual (DLSCL J0916.2+2951), gravitation, methods: analytical, methods: statistical} 

\section{Introduction}\label{sec_intro}
Merging galaxy clusters have become important astrophysical probes providing constraints on the dark matter (DM) self-interaction cross-section \citep[$\sigma_{\rm DM}$;][]{Markevitch:2004dl, Randall:2008hs, Merten:2011gu, Dawson:2012dl}, and the large-scale matter-antimatter ratio \citep{Steigman:2008dy}.
They are a suspected source of extremely energetic cosmic rays \citep{vanWeeren:2010dn}, and the merger event potentially affects the evolution of the cluster galaxies \citep[e.g.][]{Poggianti:2004ca,Hwang:2009ip,Chung:2009bz}.
All of the respective astrophysical conclusions drawn from merging clusters depend on the specific dynamic properties of a given merger.


For example, the subclass of dissociative mergers, in which the collisional cluster gas has become dissociated from the near collisionless galaxies and dark matter,  provides four ways of constraining the dark matter self-interaction cross-section \citep{Markevitch:2004dl,Randall:2008hs}.
The best constraints come from studying the mass-to-light ratios ($M/L$) of the subclusters\footnote{I define \emph{subcluster} as either one of the two colliding clusters, irrespective of mass, and I define \emph{cluster} as the whole two-subcluster system.}, and the offset between the collisionless galaxies and dark matter \citep{Markevitch:2004dl,Randall:2008hs}.
Both constraints directly depend on the merger dynamics.
First the relative collision velocity will affect the expected momentum transfer between each subcluster's dark matter particles which will in turn affect the expected dark matter mass transfer from the smaller subcluster to the larger subcluster ultimately affecting the expected mass to light ratios of the clusters \citep{Markevitch:2004dl}.
Second the expected galaxy--dark matter offset will depend on the observed time-since-collision\footnote{I define the time of collision to be the time of the first pericentric passage.} ($TSC$).
Initially the offset between the galaxies and dark matter will increase with $TSC$ (for $\sigma_{\rm DM} > 0$) as the collisionless galaxies outrun the dark matter that experienced a drag force during the collision, then at later $TSC$ the offset will decrease due to the gravitational attraction between the galaxies and dark matter halo. 
Additionally it is important to know the velocity so that dark matter candidates with velocity dependent cross-sections \citep[e.g.][]{Colin:2002ku,Vogelsberger:2012dy} can be constrained.

\begin{figure}
\epsscale{1}
\plotone{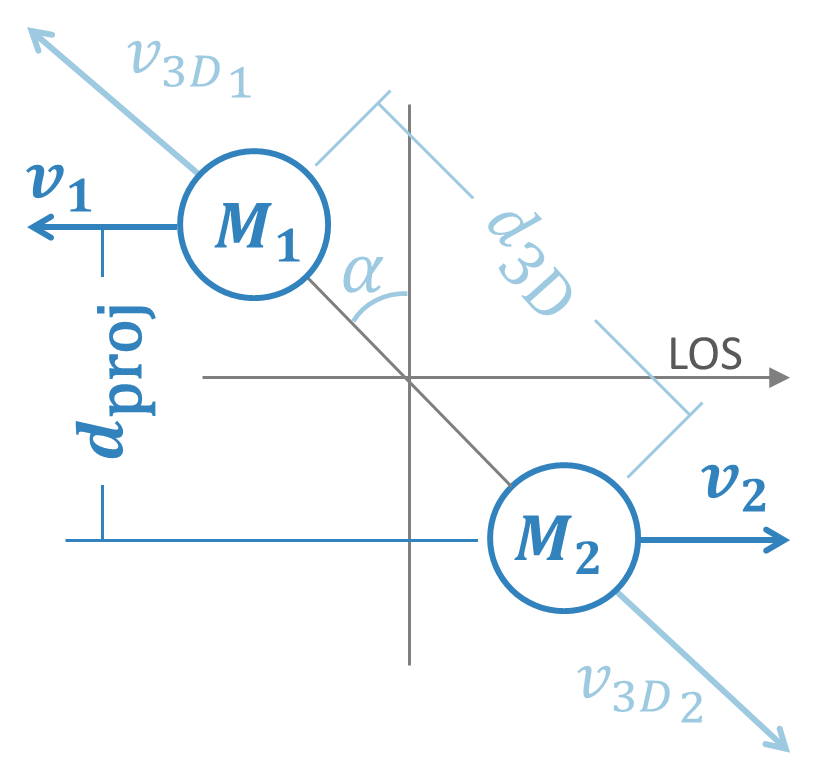}
\caption{The generic two-halo merger configuration assumed in this work.  Observable parameters are shown in dark blue, and include the mass of each halo ($M_i$), the  projected separation ($d_{\rm proj}$), and the line of sight (LOS) velocity components ($v_i$) as determined from the halo redshifts. 
The generally unknown parameters of the mergers are shown in light blue, and include the angle of the merger axis with respect to the plane of the sky ($\alpha$), and the three-dimensional separation ($d_{\rm 3D}$) and velocity components ($v_{\rm 3D_i}$).
Note that while just the outgoing scenario is shown in this figure, the method also considers the incoming scenario.
\label{fig_mergerdiagram}}
\end{figure}

However there is no way to directly observe the dynamic merger parameters of principal interest: the three-dimensional relative velocity ($v_{\rm 3D}$) and separation ($d_{\rm 3D}$) of the subclusters as a function of time, their maximum separation ($d_{\rm max}$), the period between collisions ($T$), and the time-since-collision ($TSC$).
Observations are generally limited to: the subcluster projected separation ($d_{\rm proj}$), the line-of-sight (LOS) velocity of each subcluster ($v_i$) as inferred from their redshifts, and their mass ($M_i$) or projected surface mass density profile.
In addition to the obvious inability to measure a change in the merger state, it is difficult to constrain the dynamic parameters of interest even in the observed state. 
This is due to the general inability to constrain the angle of the merger axis with respect to the plane of the sky ($\alpha$), see Figure \ref{fig_mergerdiagram}.

For the \object{Bullet Cluster} it was originally thought that estimates of the Mach number of the cluster merger through X-ray observations of the gas shock feature \citep[e.g.][]{Markevitch:2006wv} could be used to estimate $v_{\rm 3D}$, and in conjunction with measurements of the relative LOS velocities then estimate $\alpha$.
Similarly, the gas pressure differential across cold front features seen in some merging clusters have also been used to estimate the Mach number of the cluster merger \citep[e.g.][]{Vikhlinin:2003wy}. 
However, \citet{Springel:2007bg} showed that the Mach number only translates to an upper limit on $v_{\rm 3D}$, and in the case of the \object{Bullet Cluster} they showed that the Mach inferred velocity could be a factor of $\sim2$ larger than the true $v_{\rm 3D}$.
There is potential for constraining $\alpha$ using polarization measurements of radio relics \citep{Ensslin:1998tx}, which are associated with some cluster mergers \citep[e.g.\,][]{vanWeeren:2010dn} but not all \citep[e.g.\,][]{Russell:2011hn}. 
Even if for some mergers radio relics provide constraints on $\alpha$, dynamic models are still needed in order to ascertain the dynamic properties of the merger throughout time.

The two most prevalent methods for ascertaining the dynamics of observed merging systems are \emph{the timing argument} and N-body simulations.  
The timing argument is based on the solution to the equations of motion of two gravitating point masses, with the cosmological constraint that as $z \rightarrow \infty$ the separation of the two masses $d_{\rm 3D} \rightarrow 0$ \citep[for an exposition of this method see][]{Peebles:1993vp}.  The timing argument was first used by \citet{Kahn:1959ds} to study the system of the Milky Way and M31, and first applied to binary cluster systems by \citet{Beers:1982dp}.
It has recently been applied to several dissociative mergers, including the \object{Bullet Cluster} \citep{Barrena:2002dj}, \object{Abell 520} \citep{Girardi:2008gu},  \object{Abell 2163} \citep{Bourdin:2011gr}, and  \object{Abell 1758N} \citep{Boschin:2012he}.
N-body simulations of observed dissociative mergers have been limited to the \object{Bullet Cluster} (\object{1ES 0657-558}) and have come in two variants: hydrodynamic \citep{Springel:2007bg, Milosavljevic:2007hf, Mastropietro:2008fs}, and self interacting dark matter (SIDM) plus collisionless galaxy particles \citep{Randall:2008hs}.

While the timing argument method is easy to use, its inherent assumptions result in non-negligible error for dissociative systems.
Most importantly the timing argument method assumes two point mass particles; this assumption begins to break down as the two subclusters overlap and results in divergent solutions as the subclusters near collision.
Since most dissociative mergers are observed with the two-subcluster halos overlapping and dark matter constraints depend on the merger dynamics near collision, application of the traditional timing argument to dissociative mergers is limited and should be done with caution.
\citet{Nusser:2008iw} addressed this weakness of the traditional timing argument by substituting truncated \citet*[][hereafter NFW]{Navarro:1996ce} halos and numerically solving the equations of motion.
Another weakness of the timing argument is that its main constraint requires the  assumption that the subcluster masses are constant since the beginning of the universe.  
While \citet{Angus:2007em} have noted this problem with the initial conditions of N-body simulations and proposed a solution based on estimating the mass accretion histories of the clusters \citep[e.g.][]{Wechsler:2002kh}, their correction is incompatible with the timing argument method as this would add a second differential term to the equations of motion.
Finally, the large covariance between the merger parameters plus the complexity of the equations of motion makes propagation of errors in the timing argument formalism untenable.
This has resulted in a lack of certainty with timing argument results, leaving most users to run a few scenarios in an effort to roughly bound the range of possible solutions \citep[e.g.][]{Boschin:2012he}.  

N-body simulations provide the most accurate description of merger dynamics, however they are computationally expensive which results in their application being limited.
Despite eleven currently confirmed dissociative mergers\footnote{(1) Bullet Cluster \citep{Clowe:2004eq}; 
(2) A520 \citep{Mahdavi:2007ed}; 
(3) MACS J0025.4-1222 \citep{Bradac:2008gw}; 
(4) A1240 \citep{Barrena:2009to}; 
(5) ZwCL 0008.8+5215 \citep{vanWeeren:2011ko}; 
(6) A2744 \citep{Merten:2011gu}; 
(7) A2163 \citep{Okabe:2011gv}; 
(8) A1758N \citep{Ragozzine:2011jj}; 
(9) Musket Ball Cluster \citep{Dawson:2012dl}; 
(10) ACT-CL J0102−4915 \citep{Menanteau:2012bf}; 
(11) MACS J0717.5+3745 \citep{Mroczkowski:2012vs}} only the \object{Bullet Cluster} has been modeled with N-body simulations, whereas most of these have been analyzed with the timing argument method.
Existing N-body simulation strategies to ascertain the dynamic properties of mergers are incapable of keeping up with the current faster than exponential rate of discovery.
Even for the case of the \object{Bullet Cluster} the N-body analyses have been limited as far as mapping out the merger dynamic phase space allowed by  the uncertainty of the observations, with at most 15 different scenarios being run \citep{Mastropietro:2008fs}.
\citet{Gomez:2012ex} have come the closest to addressing this issue in their investigation of potential dissociative mergers (\object{A665} and \object{AS1063}) through the use of simplified scale-free numerical simulations of the mergers \citep[see][for details]{Gomez:2000kj}.  
However, they have still had to severely limit the phase space probed (fixing merger parameters such as the initial relative velocity and subcluster-subcluster mass ratio); thus admittedly this approach enables construction of plausible models, but not a thorough accounting of possible or likely models.


With these weaknesses in mind I present a new method\footnote{This method is similar to the one used by \citet{Dawson:2012dl}, although with several improvements (see \S\ref{sec_musketball}).}  for analyzing the dynamics of observed dissociative mergers.
My primary objectives are to 1) obtain a solution valid near the collision state, 2) fully estimate the covariance matrix for the merger parameters, 3) be able to analyze a dissociative merger on the order of a day using a typical desktop computer, and 4) obtain approximately 10\% accuracy; all assuming that only the most general merger observables and their uncertainty are known: mass of each subcluster, redshift of each subcluster, and projected separation of the subclusters.


In \S\ref{sec_method} I define a method for analyzing the dynamics of observed dissociative mergers.
In \S\ref{sec_sfcomp} I verify this method with existing results from a hydrodynamic N-body simulation.
In \S\ref{sec_bullet} I apply this method to the \object{Bullet Cluster} and in \S\ref{sec_musketball} I apply this method to the \object{Musket Ball Cluster} (\object{DLSCL J0816.2+2951}) and contrast its dynamics with those of the \object{Bullet Cluster}.
Finally in \S \ref{summary} I summarize my findings, discuss their implications for the constraints on dark matter and suggest other science that will benefit from the introduced method.
Throughout this paper I assume $\Omega_{\Lambda}=0.7$, $\Omega_m=0.3$, and $H=70$\,km\,s$^{-1}$\,Mpc$^{-1}$.

\section{Method}\label{sec_method}

In order to obtain a valid solution of the system dynamics near the collision state I use a model of two spherically symmetric NFW halos, rather than point masses.
I incorporate this model in a standard Monte Carlo implementation: draw randomly from the observables' probability density functions (PDF's) to generate a possible realization of the merger, use the model to calculate merger properties of interest, apply multiple priors, store these likelihood weighted results as a representative random draw of their PDF, and repeat.  
The final result is a multidimensional PDF for the dynamic parameters of the merger. 
This method agrees well with hydrodynamic simulations, \S\ref{sec_sfcomp}, and satisfies the speed and accuracy objectives outlined in the Introduction.

\subsection{Model}\label{sec_model}
The general basis of the model is a collisionless two body system with the mass of each body mutually conserved throughout the merger.
The model requires minimal input: the mass of each subcluster, the redshift of each subcluster, and the projected separation of the subclusters (along with associated uncertainties).
It assumes conservation of energy and zero angular momentum.
The model also assumes that the maximum relative velocity of the two bodies is the free-fall velocity of the system assuming their observed mass.
In the remainder of this subsection I will discuss in detail these general assumptions, their justification, and their implications. 

I model the system using two spherically symmetric NFW halos truncated at $r_{200}$\footnote{$r_{200}$ is defined as the radius of the spherical region within which the average density is 200 times the critical density at the respective redshift.}.  
By default the concentration of each halo is determined by the halo's mass via the mass-concentration scaling relation of \citet{Duffy:2008jy}.
This is not a requirement of the model though, and measured concentrations can be used, as in the case of \S\ref{sec_bc_obsprop}. 
The dynamic parameter results are relatively insensitive to the assumed concentration of the subclusters.  
Take for example the case of \S\ref{sec_sfcomp} with user specified concentrations of c$_1=1.94$ and c$_2=7.12$: if instead \citet{Duffy:2008jy} inferred concentrations c$_1=3.44$ and c$_2=2.75$ ($\sim200\%$ difference for both) are used, the difference in the estimated $v_{\rm 3D}(t_{\rm col})$ and $TSC$ are both less than 6\%.

The model assumes that the mass of each subcluster is constant and equal to the observed mass\footnote{For subcluster mass I refer to $M_{200}$, which is the mass of the individual subcluster enclosed within a radius of $r_{200}$.}.  
While this assumption is also used in the timing argument method, it is more reasonable for this method since the bulk of the results are calculated between the observed state and the collision state, typically lasting $\lesssim 1$\,Gyr.
This is about an order of magnitude shorter than the typical timescales of the timing argument method thus the new method is less susceptible to error due to neglecting growth of structure.

The model assumes that the energy of the two-halo system is conserved, and consists only of their mutual kinetic and potential energies.  
The kinetic energy of the system is $K(t) =  0.5 \mu v_{3D}(t)^2$, where $\mu$ is the reduced mass of the system and $v_{3D}(t)$ is the relative physical velocity of the two subclusters at time $t$.
The potential energy of the system is assumed to be purely gravitational and is derived in Appendix \ref{potentialsec}.
Since the model assumes zero impact parameter there is no rotational kinetic energy term.
\citet{Mastropietro:2008fs} find that a moderate impact parameter of $\sim0.1 r_{200}$ has less than a 1\% effect on the merger velocity, thus this assumption should have negligible effect for the case of dissociative mergers which must have had relatively small impact parameters in order to dissociate the bulk of their gas.

For the relative velocity of the two subclusters I apply a flat prior from zero to the free-fall velocity of the subclusters, assuming their observed mass.
This will result in an overestimate of the maximum possible relative velocity, due to the neglect of mass accretion.
It is conceivable that this prior could be tightened using the maximum relative velocities observed in cosmological N-body simulations as a function of subcluster masses and redshift. 
Another possibility for tightening the prior would be to analytically estimate the free-fall velocity accounting for mass accretion \citep[e.g.][]{Angus:2007em}.
An advantage of the Monte Carlo approach taken with this method is that additional priors can be applied as more knowledge becomes available without the need to rerun the analysis, so I opt for a conservative default approach.

The model ignores the effects of surrounding large scale structure and simply treats the two-body system.
As \citet{Nusser:2008iw} shows, a global overdense region (10 times denser than the background) engulfing the system only affects the dynamics substantially for extreme collision velocities ($\sim 4500$\,km\,s$^{-1}$).  While global overdensities may be disregarded it is not clear that the effects of nearby structures can be disregarded, e.g. as in the case three body systems.  Thus this method should be applied with caution to complex cluster mergers.

The model also ignores dynamical friction.
\citet{Farrar:2007fc} found that including dynamical friction accounted for an $\sim$10\% reduction in the inferred collision velocity of the \object{Bullet Cluster} in their analytic treatment.
This is potentially concerning since dynamical friction is inversely proportional to the relative velocity of the merger, thus it may become even more important for mergers slower than the \object{Bullet Cluser}.
However in \S \ref{sec_sfcomp} I compare the results of my method with those from a hydrodynamic N-body simulation and show that the net effect of all simplifications (including ignoring dynamical friction, tidal stripping of mass and gas mass lost during the collision) are negligible, suggesting that dynamic friction is less important than the analytic estimates of \citet{Farrar:2007fc} suggest.

\subsection{Monte Carlo Analysis}\label{sec_MCanalysis}

In this section I discuss the details of the Monte Carlo analysis workflow.
I chose to implement a Monte Carlo analysis because the high degree of correlation among the many merger dynamic parameters made traditional propagation of errors unfeasible.
A Monte Carlo analysis has the added advantage of easily enabling application of different combinations of priors ex post facto, see e.g. \S\ref{sec_addedprior}. 

The analysis begins by randomly drawing from the PDF's of the merger observables: mass of each subcluster ($M_{200_i}$), redshift of each subcluster ($z_i$), and projected separation of the subclusters ($d_{\rm proj}$).  The potential energy, $V$ (see Appendix \ref{potentialsec}), at the time of the collision is used to calculate the maximum relative velocity,
\begin{displaymath}
v_{\rm 3D_{max}} = \sqrt{-\frac{2}{\mu}V(r=0)}.
\end{displaymath}

The velocity of each subcluster relative to us is estimated from its redshift,
\begin{displaymath}
v_i = \left[\frac{(1+z_i)^2-1}{(1+z_i)^2+1}\right] c,
\end{displaymath}
where c is the speed of light.
The relative radial velocity of the subclusters is calculated from their redshifts,
\begin{displaymath}
v_{\rm rad}(t_{\rm obs}) = \frac{|v_2-v_1|}{1-\frac{v_1 v_2}{c^2}}.
\end{displaymath}

Since the angle of the merger axis with respect to the plane of the sky, $\alpha$, is unconstrained without prior knowledge of the three-dimensional relative velocity, I assume that all merger directions are equally probable.
However, projection effects result in $PDF(\alpha) = \cos(\alpha)$.
Due to symmetry it is only necessary to analyze the range $0\le\alpha\le 90$\,degrees.
I draw randomly from this PDF for each realization.
This enables the calculation of the three-dimensional relative velocity in the observed state, 
\begin{equation}
v_{\rm 3D}(t_{\rm obs}) = v_{\rm rad}(t_{\rm obs})/\sin(\alpha),
\label{eq_v3Dobs}
\end{equation}
as well as the observed three-dimensional separation of the subclusters, 
\begin{equation}
d_{\rm 3D}(t_{\rm obs}) = d_{\rm proj}/\cos(\alpha).
\label{eq_d3D}
\end{equation}

If $v_{\rm 3D}(t_{\rm obs}) > v_{\rm 3D_{max}}$, then this realization of the merger is discarded; otherwise the relative collision velocity is calculated,
\begin{equation}
v_{\rm 3D}(t_{\rm col}) = \sqrt{v_{\rm 3D}(t_{\rm obs})^2+\frac{2}{\mu}\left[V(t_{\rm obs})-V(t_{\rm col})\right]}.
\label{eq_v3Dcol}
\end{equation}
Similarly if $v_{\rm 3D}(t_{\rm col}) >v_{\rm 3D_{max}}$, then this realization is discarded.

The change in time, $\Delta t$, between two separations is given by
\begin{equation}
\Delta t = \int_{r_1}^{r_2} \frac{dr}{\sqrt{\frac{2}{\mu}(E-V(r))}}.
\label{eq_deltat}
\end{equation}
I define the time-since-collision ($TSC$) as the time it takes the subclusters to traverse from zero separation to their physical separation in the observed state, $d_{\rm 3D}$. 
Because there is a potential degeneracy in whether the subclusters are ``outgoing'' (approaching the apoapsis after collision)  or ``incoming'' (on a return trajectory after colliding and reaching the apoapsis); I solve for both of these cases, $TSC_0$ and $TSC_1$ respectively.
In determining $TSC_1$ it is useful to define the period, $T$, of the system.  I define $T$ to be the time between collisions,
\begin{displaymath}
T  = 2 \int_{0}^{d_{\rm max}} \frac{dr}{\sqrt{\frac{2}{\mu}(E-V(r))}},
\end{displaymath}
where $d_{\rm max}$ is the distance from zero separation to the apoapsis, when $E=V$.
Thus,
\begin{displaymath}
TSC_1 = T - TSC_0.
\end{displaymath}
During the Monte Carlo analysis any realizations with $TSC_0$ greater than the age of the Universe at the cluster redshift are discarded.
A similar flat prior is applied when calculating the statistics of $TSC_1$.
To this regard some insight into the likelihood of the system being in an ``outgoing'' or ``incoming'' state can be gained by calculating the fraction of realizations with $TSC_1$ less than the age of the Universe at the cluster redshift.
Conceivably these temporal priors could be strengthened, requiring that the time to first collision ($T$) plus the respective $TSC$ be less than the age of the Universe at the cluster redshift, in a fashion similar to the timing argument.
However, as with the timing argument model, the model of \S\ref{sec_model} becomes less valid over time-scales approaching the age of the Universe.
Thus I use the more conservative prior by default.

Since the majority of the merger time is spent at large separations, due to lower relative velocities, observations of the system are more likely near apoapsis than near the collision.  Thus the probability of each realization is convolved with the prior 
\begin{equation}
{\rm PDF}(TSC_0) = 2 \frac{TSC_0}{T}.
\label{eq_timeprior}
\end{equation}
There are likely selection effects which complicate this PDF, since it can be imagined that the X-ray luminosity is greatest near the time of the collision \citep[see e.g.][]{Randall:2002kk}.
However this information is rarely if ever known, thus it is not included by default.  In \S\ref{sec_addedprior} I show how additional temporal priors, based on similar effects, may be effectively applied to the results of the analysis ex post facto.

The end result of this method is a 13 dimensional posterior PDF of an array of cluster merger parameters, see for example Appendix \ref{sec_bcresults}.
Finally to compact the results I use the biweight-statistic (generally more robust and less sensitive to abnormally tailed distributions than the median or mean) and bias-corrected percent confidence limits \citep{Beers:1990kg} applied to the marginalized parameter distributions of the valid realizations, see for example Table \ref{bulletresultparam}.

\subsection{Comparison with Hydrodynamic Simulations} \label{sec_sfcomp}

\begin{figure}
\plotone{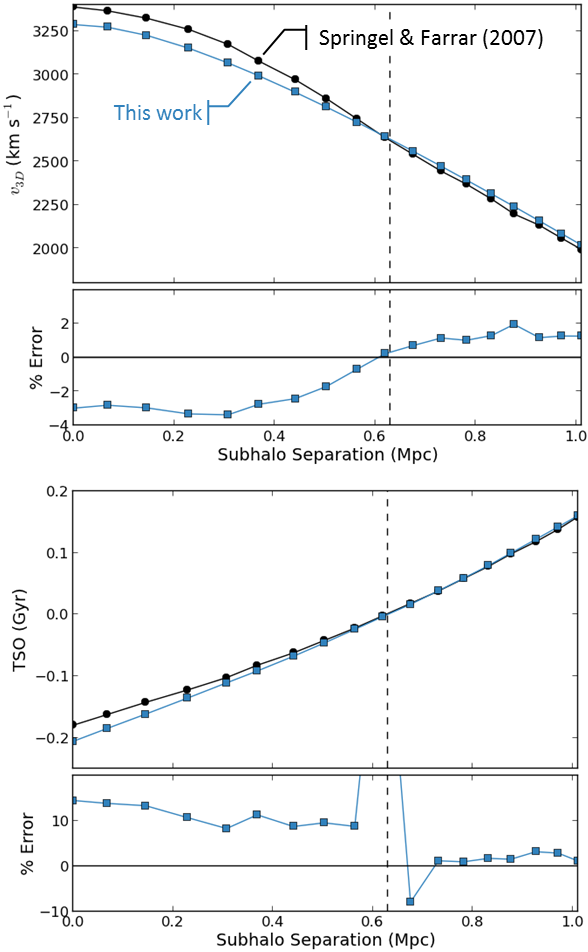}
\caption{
Comparison of the model results (blue boxes) with the hydrodynamic simulation results of \citet[black circles;][]{Springel:2007bg} for the Bullet Cluster.  
The top figure is a comparison of the velocity of the ``bullet'' relative to the ``main'' subcluster, with the subhalo separation (i.e. the three-dimensional separation of the ``main'' and ``bullet'' subclusters) as the independent variable.  
The bottom figure is a comparison of the time-since-observed state (TSO), where the ``observed'' state (dashed line) is defined by \citet{Springel:2007bg} as the time step in their simulation when the gas properties most closely match the observed Bullet Cluster gas properties.
Times prior(post) to the observed state have negative(positive) values.  
The percent error in each case is calculated assuming the \citet{Springel:2007bg} results as truth.  
While the model results are biased, the bias appears stable and is roughly an order of magnitude smaller than the typical random error in the parameter estimates (see for example Table \ref{bulletresultparam}).  
Note that the TSO percent error calculation understandably diverges near the arbitrary choice of time equal zero. 
The \citet{Springel:2007bg} results are read directly from their Figure 4.
\label{fig_SFcomparison}}
\end{figure}

For the purposes of checking the physical assumptions of the model I reanalyze the \citet{Springel:2007bg} model of the \object{Bullet Cluster}, comparing my dynamic parameter estimates with their hydrodynamic N-body simulation based estimates.
For this analysis I run just their single case through the model (i.e.\,I do not perform a Monte Carlo analysis).
They represent the ``main'' and ``bullet'' subclusters as NFW halos with  M$_{200_1} = 1.5\times 10^{15} $\,M$_\sun$, c$_1=1.94$, M$_{200_2} = 1.5\times 10^{14}$\,M$_\sun$, and c$_2=7.12$, respectively.
They note that the gas properties of their simulation most closely match the observed \object{Bullet Cluster} gas properties for the time step corresponding to a subcluster separation of $d_{3D}=625$\,kpc and  relative velocity of $v_{\rm 3D}(t_{\rm obs})=2630$\,km\,s$^{-1}$.
I define this as the ``observed'' state (dashed line in Figure \ref{fig_SFcomparison}) and use the model discussed in \S \ref{sec_model} to extrapolate values of the relative subcluster velocities ($v_{\rm 3D}$) and time-since-observed state (TSO) before and after the observed state (left and right of the dashed line in Figure \ref{fig_SFcomparison}, respectively).
The \citet{Springel:2007bg} simulation results (black circles) for these parameters are read directly from their Figure 4.

I compare the model results (blue boxes) with the \citet{Springel:2007bg} simulation results, and assume their results as truth when calculating the percent error, see Figure \ref{fig_SFcomparison}.
There is better than 4\% agreement between $v_{\rm 3D}$ and 14\% agreement between the TSO.
While the model results are biased, the bias appears stable and is roughly an order of magnitude smaller than the typical random error in the parameter estimates (see for example Table \ref{bulletresultparam}).
Given the stability of the bias it is conceivable that it could be corrected in the model results.
However, to have any confidence in this bias correction the model results should be compared with a range of merger scenarios, which is beyond the scope of this current work.
Note that the better agreement between the velocity estimates than between the TSO estimates is to be expected since the velocity calculation (essentially Equation \ref{eq_v3Dcol}) comes from simply comparing the observed and another state of the merger whereas the TSO calculation (Equation \ref{eq_deltat}) requires integration between these two states.
The results of this comparative study essentially validate many of the simplifying assumptions of the model (conservation of energy, and ignoring the affects of dynamical friction, tidal stripping of dark matter and gas during the collision).

As an aside it should be noted that for this comparison I use the \citet{Springel:2007bg} NFW halo parameters that represent the state of the halos prior to collision.
Ideally I should use the NFW parameters representative of the state of the halos at $t_{\rm obs}$, however these properties were not reported in their paper.
From Figure 5 of \citet{Springel:2007bg} some insight into the time variability of the halo parameters can be gained.
Since the depth each halo's gravitational potential at $\sim r_{200}$ does not change appreciably throughout the merger, it can be inferred that M$_{200}$ of each halo does not change. 
However, the gravitational potential near the center of each halo deepens by $\sim 25\%$ during and after the collision.
This can be interpreted as the concentration of each halo increasing.
Thus for the comparison of my model with the \citet{Springel:2007bg} hydrodynamic simulation to be more appropriate I should have used halos with larger concentrations.
Doing so actually brings my model results more in-line with the simulation results.
If for example I increase the concentration of the ``bullet'' halo from 1.94 to 3 and the concentration of the ``main'' halo from 7.12 to 8, then the percent error for the relative velocity of the halos reduces to $\lesssim 1\%$ and the percent error for the TSO reduces to $\sim 10\%$. 
Thus the comparative results of Figure \ref{fig_SFcomparison} should be considered conservative with respect to the variability of the halo properties throughout the merger.

%

\section{Bullet Cluster Dynamics}\label{sec_bullet}

The \object{Bullet Cluster} is the prime candidate for first application of the method as it is one of the best studied dissociative mergers.
It has a wealth of observational data necessary for input to the model, as discussed in \S\ref{sec_bc_obsprop}, plus supplementary data which enables additional posterior priors, as discussed in \S\ref{sec_addedprior}.

\subsection{Bullet Cluster Observed System Properties}\label{sec_bc_obsprop}

I summarize the observed \object{Bullet Cluster} parameters used as input to my analysis in Table \ref{bulletinputparam}.  
The full PDF's of these input parameters have not been published so I simply assume Gaussian distributions.
I refer to the main subcluster as halo 1 and the ``bullet'' subcluster as halo 2.  
For the mass and concentration of each subcluster I use the most recently reported estimates from \citet{Springel:2007bg}, based upon strong and weak lensing estimates \citep{Bradac:2006be}.
However, they do not present errors for these quantities so for the mass I estimate the  1--$\sigma$ errors to be 10\% of the mass, since this is approximately the magnitude of the error reported by \citet{Bradac:2006be} for M($<250$\,kpc).
There is no published estimate for the uncertainty of the concentrations of the NFW model fits, $c_i$, so I simply assume the concentrations to be known quantities (as noted in \S\ref{sec_model} the results are relatively insensitive to the assumed concentrations). 
The redshifts of the main and ``bullet'' subclusters are estimated from 71 and 7 spectroscopic members, respectively \citep{Barrena:2002dj}.
The projected separation of the mass peaks is determined from strong and weak lensing measurements \citep{Bradac:2006be}, and is essentially the same as the separation of the subclusters' galaxy centroids.
For each Monte Carlo realization individual values are drawn randomly from each of these assumed Gaussian distributions.

\begin{deluxetable}{lcccc}
\tablewidth{0pt}
\tablecaption{Bullet Cluster parameter input\label{bulletinputparam}}

\tablehead{
\colhead{Parameter}     & \colhead{Units}  & \colhead{$\mu$}  & \colhead{$\sigma$}	&	\colhead{Ref.}
}
\startdata
M$_{200_1}$	&	$10^{14}$\,M$_\sun$	&	15	&	1.5\tablenotemark{a}	&	1	\\
c$_1$	&		&	7.2	&	\nodata\tablenotemark{b}	&	1	\\
M$_{200_2}$	&	$10^{14}$\,M$_\sun$	&	1.5	&	0.15\tablenotemark{a}	&	1	\\
c$_2$	&		&	2.0	&	\nodata\tablenotemark{b}	&	1	\\
z$_1$	&		&	0.29560	&	0.00023	&	2	\\
z$_2$	&		&	0.29826	&	0.00014	&	2	\\
$d_{\rm proj}$		&	kpc	&	720	&	25	&	3	\\
\enddata
\tablecomments{A Gaussian distribution with mean, $\mu$, and standard deviation, $\sigma$, is assumed for all parameters with quoted respective values.  The mass, M$_{200}$, and concentration, c, are the defining properties of assumed spherically symmetric NFW halos.}
\tablenotetext{a}{Estimated to be 10\%, based one the error magnitude of M($<250$\,kpc) reported in \citet{Bradac:2006be}.}
\tablenotetext{b}{No errors were presented in the reference.  A single concentration value was used for all Monte Carlo realizations.}
\tablerefs{
(1) Springel \& Farrar 2007; (2) Barrena et al. 2002; (3) Brada{\v c} et al. 2006.}
\end{deluxetable}

\subsection{Bullet Cluster System Dynamics Results} \label{bulletdynamics}

I first analyze the \object{Bullet Cluster} with the Monte Carlo analysis method and default priors discussed in \S\ref{sec_method}, highlighting the complexity of merger dynamics and the inappropriateness of analyzing a small sample of select merger scenarios.
In \S\ref{sec_addedprior} I incorporate  additional constraints provided by the observed strong X-ray shock front plus boosted temperature and luminosity.
I discuss this prior information and apply it ex post facto to the default prior results of \S\ref{sec_defaultprior}.

I perform the analysis with 2,000,000 Monte Carlo realizations.  
Parameter estimates converge to better than a fraction of a percent with only 20,000 realizations ($\sim6$\,CPU\,hours).
I run a factor of a hundred more since it was computationally inexpensive and it provides a data sample to which I can apply any number of conceivable posterior PDF's and still maintain sub-percent statistical accuracy.

\begin{figure}
\epsscale{1.15}
\plotone{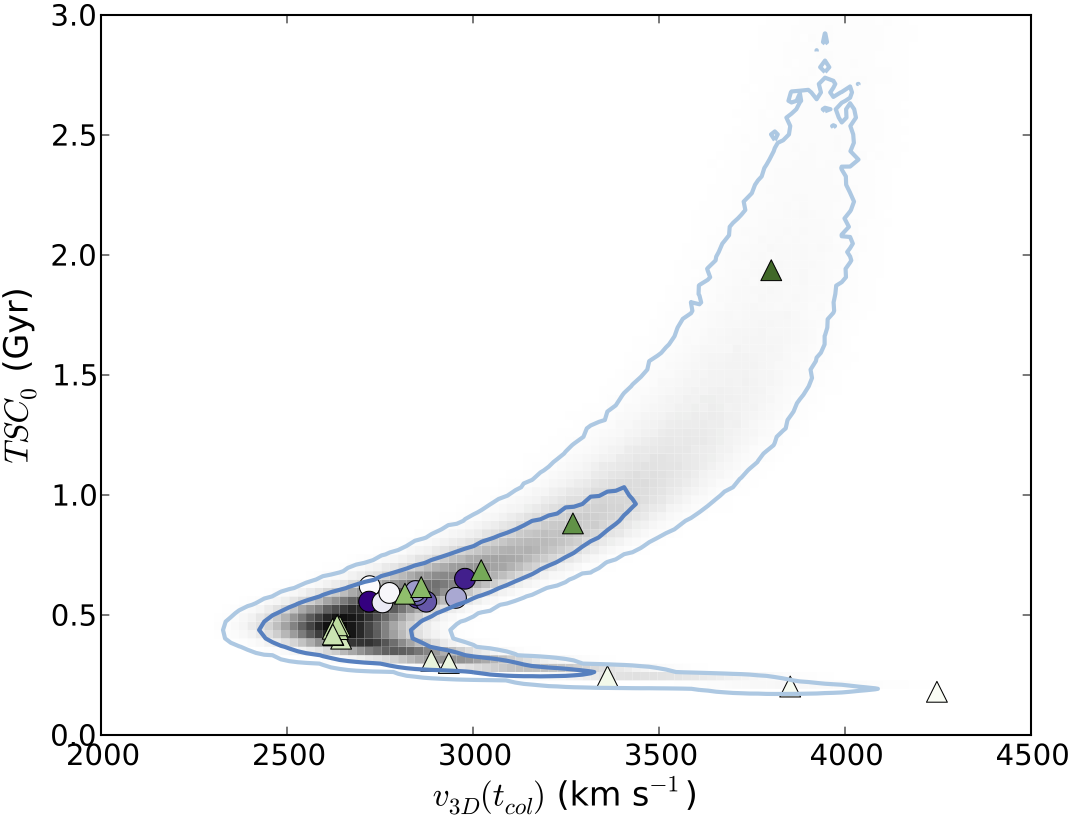}
\caption{
The posterior of the Bullet Cluster's time-since-collision $TSC_0$ and $v_{\rm 3D}(t_{\rm col})$ parameters is shown in grayscale with dark and light blue contours representing 68\% and 95\% confidence, respectively.
The green-scale triangles are from a subsample of the Monte Carlo population, which jointly satisfies the requirement of being drawn from $\pm 0.01 \sigma$ of the mean of each of the input parameters, i.e. the ``most likely'' values for the input parameters. 
Despite representing the most probable input parameter values there is considerable spread in the inferred output parameters, with the subsample clearly tracing the ridge of the distribution. 
The saturation of the triangles increases with increasing $\alpha$, from 10--86\,degrees.
The purple-scale circles are from a subsample near the bi-weight location of $\alpha=50\pm0.0002$\,degrees, with the saturation of the circles increasing with increasing M$_{200_1}$.
While the length of the distribution is predominantly caused by uncertainty in $\alpha$ the width is predominantly caused by uncertainty in the input parameters.
Despite the Bullet Cluster being one of the best measured dissociative mergers there is still considerable and complex uncertainty in its merger parameters, predominantly due to uncertainty in $\alpha$. 
\label{fig_bcTSC}}
\end{figure}

\subsubsection{Default Priors}\label{sec_defaultprior}

The main results of this analysis are that: 1) there is a great degree of covariance between the geometry, velocity, and time parameters of the merger, and 2) models of the system which disregard the uncertainty of $\alpha$ will catastrophically fail to capture the true uncertainty in the dynamic parameters.

The two-dimensional PDF of Figure \ref{fig_bcTSC} exemplifies the complexity of the covariance between the various merger parameters\footnote{Similar degrees of complex covariance exist for the other geometry, velocity and time parameters, see e.g.\,the results array in Appendix \ref{sec_bcresults}.}.
The shape of the PDF is most easily understood in terms of the parameters' dependence on $\alpha$.
This dependence is illustrated by the green-scale triangles that represent a subsample of the Monte Carlo population,  which jointly satisfies the requirement of being drawn from $\pm 0.13 \sigma$ of the mean of each of the input parameters, i.e. the ``most likely'' values for the input parameters.
The saturation of the triangles increases with increasing $\alpha$, from 10--86\,degrees, clearly showing a monotonically increasing relationship with $TSC_0$ (see also Figure \ref{fig_bc_geovt}).
For small $\alpha$ (light green triangles), Equation \ref{eq_v3Dobs} states that $v_{\rm 3D}(t_{\rm obs})$ must be large thus $v_{\rm 3D}(t_{\rm col})$ must also be large, and since Equation \ref{eq_d3D} states that $d_{\rm 3D}(t_{\rm obs})$ approaches the minimum possible observed separation, $d_{\rm proj}$, the $TSC_0$ must approach a minimum.
Conversely for large $\alpha$ (dark green triangles), $d_{\rm 3D}(t_{\rm obs})$ becomes large increasing the time required to reach the observed state, and despite $v_{\rm 3D}(t_{\rm obs})$ approaching the minimum $v_{\rm rad}(t_{\rm obs})$ the collision velocity must increase for the subclusters to have been able to reach the larger $d_{\rm 3D}(t_{\rm obs})$.   

The bulk of the uncertainty in the geometry, velocity and time parameters is due to the uncertainty of $\alpha$.
This is exemplified by the fact that the green-scale triangles in Figure \ref{fig_bcTSC} closely trace the extent of the ridge line of the two-dimensional distribution (i.e. span the bulk of the uncertainty).
Conversely the ``width'' of the distribution is predominantly due to uncertainty in the input parameters.
This is exemplified by the purple circles of Figure \ref{fig_bcTSC}, which are for a near constant $\alpha$ yet randomly sample the M$_{200_1}$ distribution. 
The saturation of the circles increase with increasing mass.

The inability to directly measure $\alpha$, coupled with its strong degree of correlation with the other dynamic parameters, makes it the dominant source of uncertainty. 
While it was originally believed that the three-dimensional merger velocity as inferred from the X-ray shock feature could be coupled with the redshift determined radial velocity to measure $\alpha$, \citet{Springel:2007bg} showed that the X-ray shock inferred velocity significantly overestimates the true three-dimensional merger velocity.
So at best this information can weakly constrain $\alpha$, and in the case of the \object{Bullet Cluster} the X-ray shock inferred velocity is significantly greater than the free-fall velocity, $v_{\rm 3D_{max}}$, thus it provides no additional constraining power.
In \S\ref{sec_suggestions} I discuss how the results of this method can be used in conjunction with N-body simulations to limit the computational impact of accounting for the uncertainty in $\alpha$.

\begin{deluxetable*}{lc|ccc|ccc}
\tablewidth{0pt}
\tablecaption{Bullet Cluster parameter estimates\label{bulletresultparam}}
\tablehead{																							
Parameter     & Units  & 
\multicolumn{3}{|c|}{Default Priors}  & \multicolumn{3}{|c}{Default + Added Temporal Priors}
\\
     &   & 
\colhead{Location\tablenotemark{a}}  & \colhead{68\% LCL--UCL\tablenotemark{b}} & \colhead{95\% LCL--UCL\tablenotemark{b}}	& 
\colhead{Location\tablenotemark{a}}  & \colhead{68\% LCL--UCL\tablenotemark{b}} & \colhead{95\% LCL--UCL\tablenotemark{b}}	
}					
\startdata																							
M$_{200_1}$	&	$10^{14}$\,M$_\sun$	&	
15.0	&	13.5	--	16.6	&	12.1	--	18.1	&	
15.2	&	13.6	--	16.6	&	12.2	--	18.1	\\
M$_{200_2}$	&	$10^{14}$\,M$_\sun$	&	
1.5	&	1.4	--	1.6	&	1.2	--	1.8	&	
1.5	&	1.4	--	1.7	&	1.2	--	1.8	\\
$z_1$	&		&	
0.2956	&	0.2954	--0.2958	&	0.2951 --0.2961	&	
0.2956	&	0.2954	--0.2958	&	0.2951 --0.2961	\\
$z_2$	&		&	
0.2983	&	0.2981	--	0.2984	&	0.2980	--	0.2985	&	
0.2983	&	0.2981	--	0.2984	&	0.2980	--	0.2985	\\
$d_{\rm proj}$	&	Mpc	&	
0.72	&	0.69	--	0.76	&	0.65	--	0.80	&	
0.72	&	0.68	--	0.75	&	0.64	--	0.79	\\
$\alpha$	&	degree	&	
50	&	27	--	73	&	15	--	84	&	
24	&	16	--	38	&	11	--	53	\\
$d_{\rm 3D}$	&	Mpc	&	
1.1	&	0.8	--	2.6	&	0.7	--	7.1	&	
0.8	&	0.7	--	0.9	&	0.7	--	1.2	\\
$d_{\rm max}$	&	Mpc	&	
1.3	&	1.1	--	2.5	&	1.0	--	6.4	&	
1.2	&	1.0	--	1.7	&	1.0	--	3.1	\\
$v_{\rm 3D}(t_{\rm obs})$	&	km\,s$^{-1}$	&	
820	&	640	--	1500	&	550	--	2500	&	
1600	&	1100	--	2500	&	790	--	3200	\\
$v_{\rm 3D}(t_{\rm col})$	&	km\,s$^{-1}$	&	
3000	&	2700	--	3800	&	2500	--	4200	&	
2800	&	2600	--	3300	&	2500	--	3800	\\
$TSC_0$	&	Gyr	&	
0.6	&	0.3	--	1.1	&	0.2	--	3.9	&	
0.4	&	0.3	--	0.5	&	0.2	--	0.6	\\
$TSC_1$\tablenotemark{c} &	Gyr	&	
1.2	&	1.0	--	2.4	&	0.9	--	8.2	&	
1.3	&	1.0	--	2.0	&	0.9	--	4.6	\\
$T$	&	Gyr	&	
1.8	&	1.5	--	3.2	&	1.4	--	8.1	&	
1.6	&	1.4	--	2.3	&	1.3	--	4.8	\\
\enddata																							
\tablenotetext{a}{Biweight-statistic location \citep[see e.g.][]{Beers:1990kg}.}																							
\tablenotetext{b}{Bias-corrected lower and upper confidence limits, LCL and UCL respectively \citep[see e.g.][]{Beers:1990kg}.}
\tablenotetext{c}{For the case of the Default + Added Temporal Prior, none of the realizations have a valid $TSC_1$, meaning that the Bullet Cluster is being observed in the ``outgoing'' state, as discussed in \S\ref{sec_addedprior}.}
\end{deluxetable*}	

\subsubsection{Added Temporal Prior}\label{sec_addedprior}

One of the advantages of this Monte Carlo method is that additional constraints are easily incorporated ex post facto.
An example of such constraints in the case of the \object{Bullet Cluster} is the observed X-ray shock front and factor of 2.4 greater X-ray estimated mass to lensing estimated mass \citep{Markevitch:2006wv}, due to merger related X-ray temperature and luminosity boost.  
Hydrodynamic simulations of merging clusters \citep[e.g.][]{Ricker:2001ju,Randall:2002kk} suggest that such transient effects last of order the X-ray sound crossing time.
Since simulations show negligible difference between the time scales of the two I chose to construct a prior based on the observed temperature boost.
\citet{Randall:2002kk} find that the full-width-half-max (FWHM) duration of the temperature boost is $\sim 0.4 t_{\rm sc}$ with the entire boost duration being $\sim 1.4 t_{\rm sc}$, where $t_{\rm sc}$ is the sound crossing time of the more massive of the two subclusters.
The peak of this boost roughly coincides with the time of the \emph{collision}, as defined in \S\ref{sec_intro}.
Given the M$_{200_1}=15 \times 10^{14}$\,M$_\sun$ and temperature T$_{\rm X}=14$\,keV of the ``main'' subcluster \citep{Markevitch:2006wv}, the $t_{\rm sc}=1$\,Gyr.
I construct a sigmoid function for the $TSC$ prior PDF based on the observed temperature boost, 
\begin{displaymath}
{\rm PDF}(TSC) = \frac{1}{2}\left[1-\tanh\left(\frac{TSC-0.5 a}{0.25 b} \right)\right],
\end{displaymath}
where $a$ is the FWHM of the duration of the temperature boost and $b$ is the entire boost duration.
I chose a sigmoid function over a simple step function since the temperature boost predicted by \citet{Randall:2002kk} does not end abruptly.
This prior is coupled with the previously discussed $TSC$ prior (Equation \ref{eq_timeprior}).

\begin{figure}
\epsscale{1.15}
\plotone{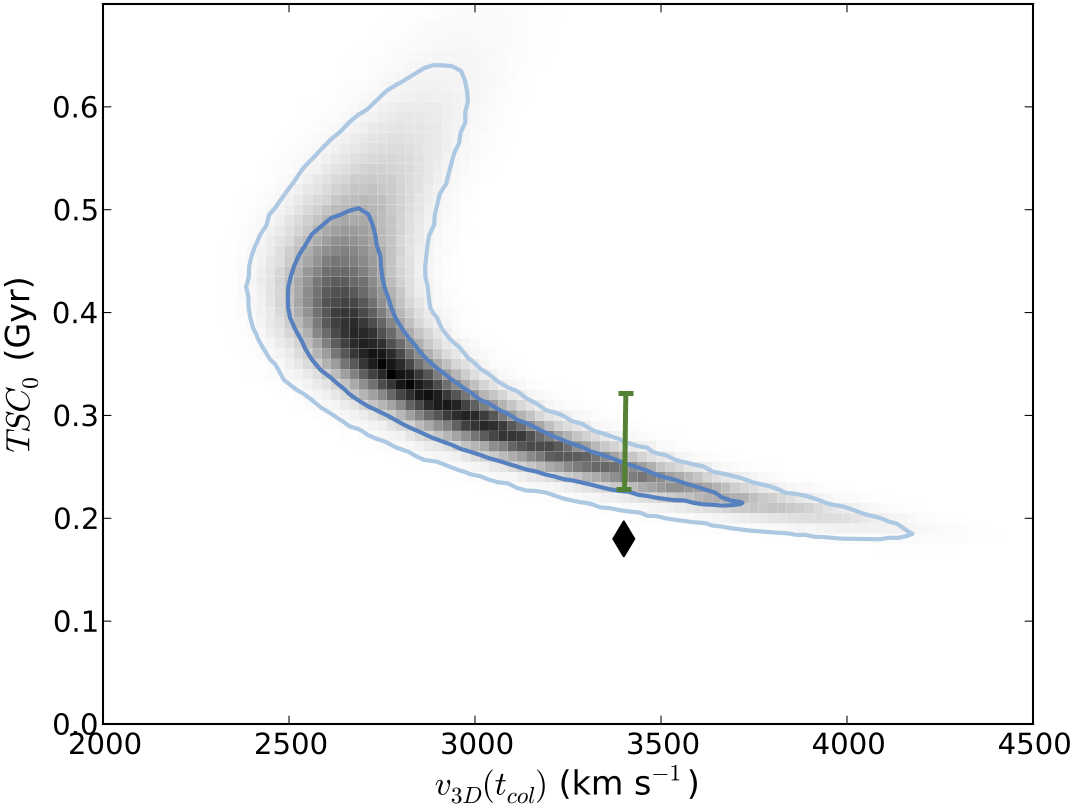}
\caption{
The posterior of the Bullet Cluster's $TSC_0$ and $v_{\rm 3D}(t_{\rm col})$ parameters after application of an additional temporal prior based on X-ray observations of the Bullet Cluster (grayscale).  
Dark and light blue contours representing 68\% and 95\% confidence, respectively.
The added temporal prior significantly improves the constraint on the merger parameters (compare with Figure \ref{fig_bcTSC}).
The black diamond represents the Springel \& Farrar (2007) hydrodynamic simulation result for their defined ``observed state'', whose X-ray properties best match the observed X-ray properties; $d_{\rm 3D}$=625\,kpc for this state.
The green bar shows their result for $d_{\rm 3D}$ between 700 to 900\,kpc, which is more in line with the observed $d_{\rm proj} = 720\pm 25$\,kpc (assuming $0<\alpha<35$\,degrees).
\label{fig_bcTSC_added}}
\end{figure}

Application of this prior significantly improves the uncertainty in $TSC_0$ (180\% to 67\%)  and $v_{\rm 3D}(t_{\rm col})$ (28\% to 19\%), compare Figure \ref{fig_bcTSC} with Figure \ref{fig_bcTSC_added}.
It essentially removes the possibility of a $TSC > 0.6$\,Gyr.
As expected from the $\alpha$ dependence shown by the green triangles in Figure \ref{fig_bcTSC} this prior also reduces the likelihood of $\alpha \gtrsim 50$\,degrees, which in turn affects both the location and uncertainty of $d_{\rm3D}$ and $v_{\rm 3D}(t_{\rm obs})$, see Table \ref{bulletresultparam}. 
The remaining parameter estimates are predominantly unaffected by the prior, with only a few having their confidence limits affected as the result of their high end low probability tails being down weighted.
Additionally there is now essentially zero probability that the bullet subcluster has reached the apoapsis and is on a return trajectory, since the 95\% lower confidence limit of $TSC_1$ is 0.9\,Gyr (see Table \ref{bulletresultparam}) and the prior essentially removes the possibility of a $TSC > 0.6$\,Gyr.
I present the full results array, which includes the default analysis prior combined with this temperature boost prior, in Appendix \ref{sec_bcresults} and the compact parameter estimates in Table \ref{bulletresultparam}.

According to this analysis the \citet{Springel:2007bg} results stated in \S\ref{sec_sfcomp} seem unlikely (see the black diamond of Figure \ref{fig_bcTSC_added}), however this is simply due to their definition of the ``observed state''.
They define the observed state to be when their simulated X-ray properties most closely match the observed X-ray properties, yet the separation of the halos in this state is only $d_{\rm 3D}=625$\,kpc; this is less than the observed $d_{\rm proj} = 720\pm 25$\,kpc \citep{Bradac:2006be}.
If we instead consider their estimate of $TSC_0$ for $d_{\rm 3D}$ between 700 to 900\,kpc (corresponding to  $d_{\rm proj} = 700$ and $0<\alpha<35$\,degrees), then $0.24<TSC_0<0.33$\,Gyr (see green bar of Figure \ref{fig_bcTSC_added}).
This brings their result in line with the results of this method, as expected by the agreement presented in \S\ref{sec_sfcomp}.
Note that the general conclusion of \citet{Springel:2007bg}, that the  shock speed greatly overestimates the actual relative speed of the subclusters, remains valid regardless of which ``observed state'' is used.

\section{Musket Ball Cluster Dynamics}\label{sec_musketball}

I also apply the method to the \object{Musket Ball Cluster}, with the objective of updating an existing analysis and comparing this system with the \object{Bullet Cluster}.  
A preliminary analysis of the system dynamics using a similar method \citep{Dawson:2012dl} suggested that the \object{Musket Ball Cluster} merger is $\sim$3--5 times further progressed than other confirmed dissociative mergers.
However, that analysis treated the two merging subclusters as uniform density spheres   and also failed to account for the temporal phase-space PDF (Equation \ref{eq_timeprior}).
Additionally the claim that the \object{Musket Ball Cluster} is both slower and further progressed than the \object{Bullet Cluster} was based on comparing the Musket Ball's $TSC_0$--$v_{\rm 3D}(t_{\rm col})$ PDF with that of the single point \citet{Springel:2007bg} estimate.
As noted in \S\ref{sec_defaultprior} there is a large area of parameter space that the \citet{Springel:2007bg} result fails to represent.

Similar to my analysis of the \object{Bullet Cluster} I perform the analysis with 2,000,000 Monte Carlo realizations.
Parameter estimates converge to better than a fraction of a percent with only 20,000 realizations.

\subsection{Musket Ball Observed System Properties}

\begin{figure}
\epsscale{1}
\plotone{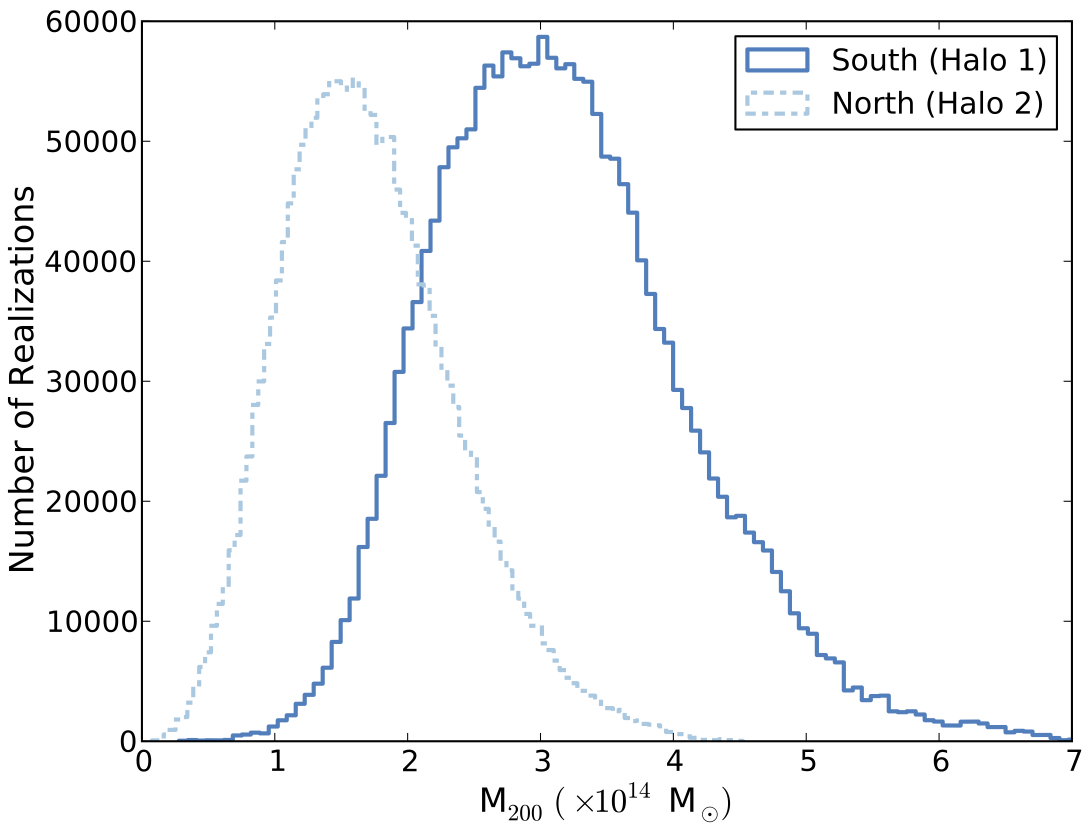}
\caption{Weak lensing mass PDF's of the Musket Ball subclusters \citep{Dawson:2012dl}. 
\label{musket_massinput}}
\end{figure}

\begin{figure}
\epsscale{1}
\plotone{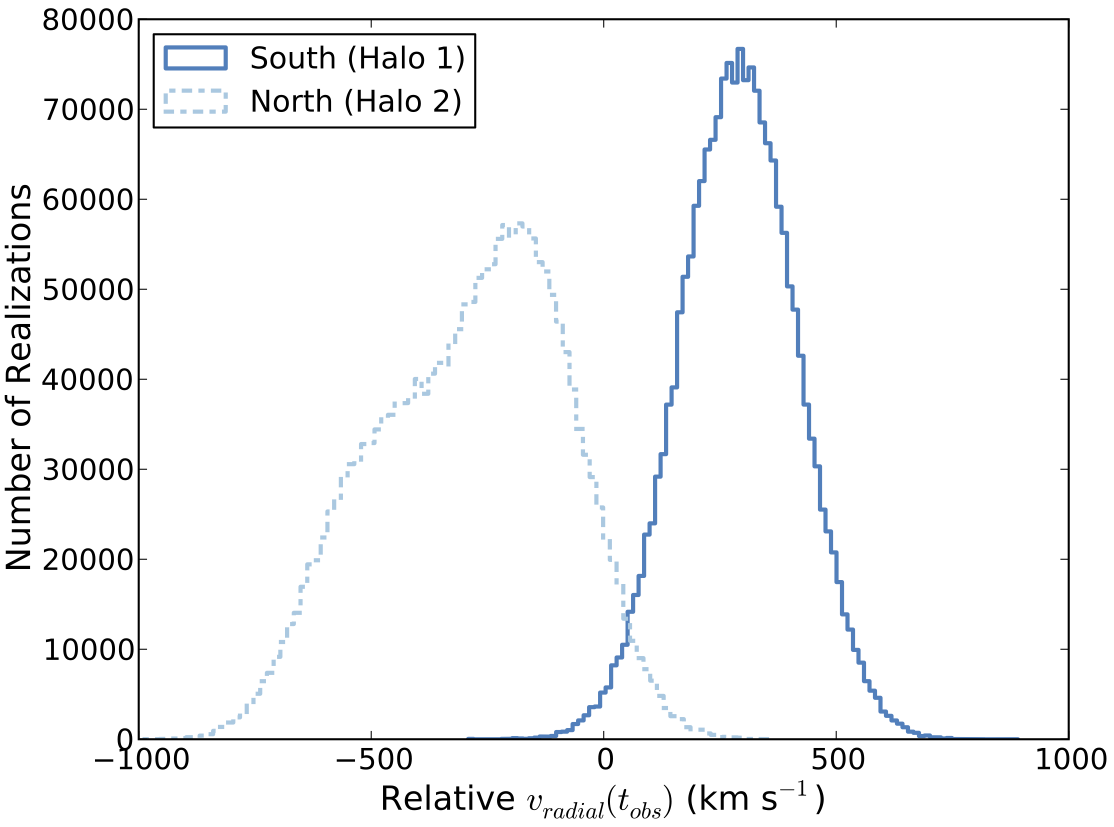}
\caption{Relative radial subcluster velocity PDF's inferred from spectroscopic redshifts the Musket Ball Cluster galaxies  \citep{Dawson:2012dl}. 
\label{musket_vinput}}
\end{figure}

\begin{figure}
\epsscale{1}
\plotone{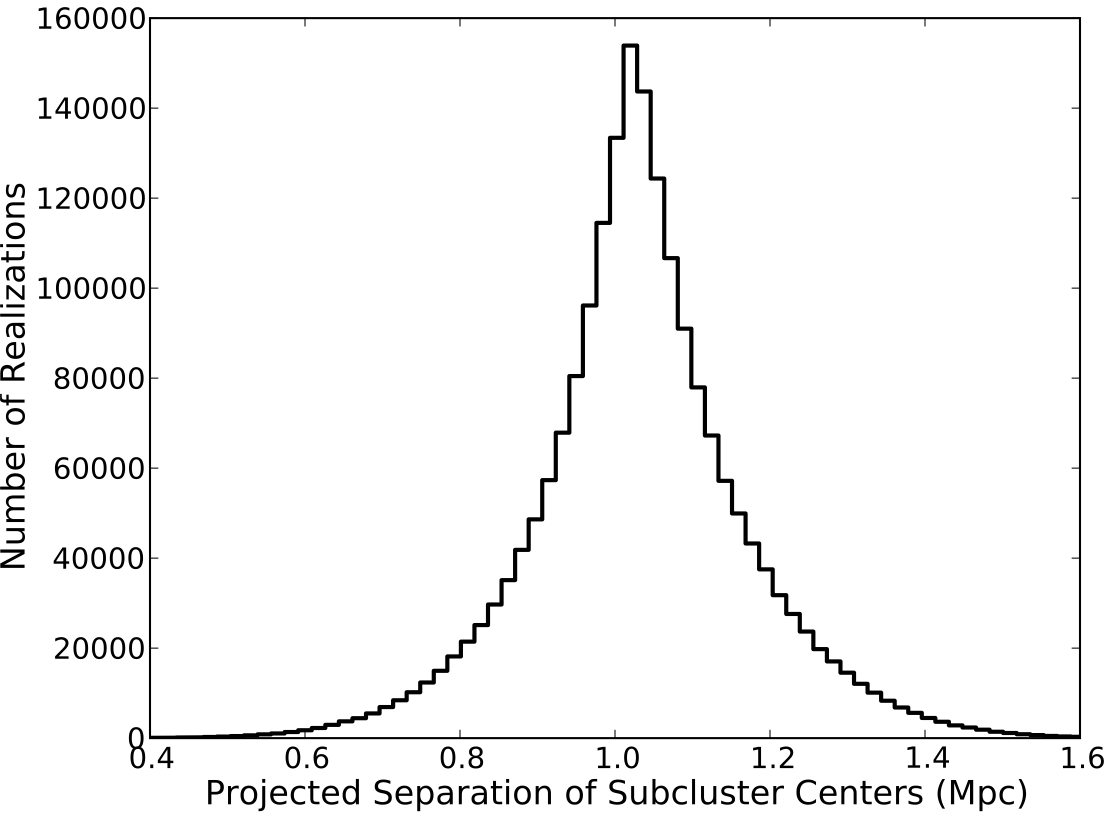}
\caption{Projected separation PDF of the Musket Ball subcluster galaxy density centroids \citep{Dawson:2012dl}. 
\label{musket_projinput}}
\end{figure}

I show the observed \object{Musket Ball Cluster} parameter PDF's in Figures \ref{musket_massinput}--\ref{musket_projinput}, each the result of analyses presented by \citet{Dawson:2012dl}.  
I refer to their ``south'' subcluster as halo 1 and ``north'' subcluster as halo 2. 
The mass PDF's, Figure \ref{musket_massinput}, are the result of an MCMC analysis where NFW halos were simultaneously fit to the weak lensing signal.
The relative velocity distributions, Figure \ref{musket_vinput}, are the result of a bootstrap error analysis \citep{Beers:1990kg} of the 38 and 35 spectroscopic members of the north and south subclusters, respectively.
The projected subcluster separation distribution, Figure \ref{musket_projinput}, is the result of a bootstrap error analysis of the recursively estimated subclusters' galaxy number density centroids \citep[see e.g.][for a description of this method]{Randall:2008hs}.
For each Monte Carlo realization individual values are drawn randomly from each of these distributions. 

\subsection{Musket Ball System Dynamics Results}

\begin{figure}
\epsscale{1}
\plotone{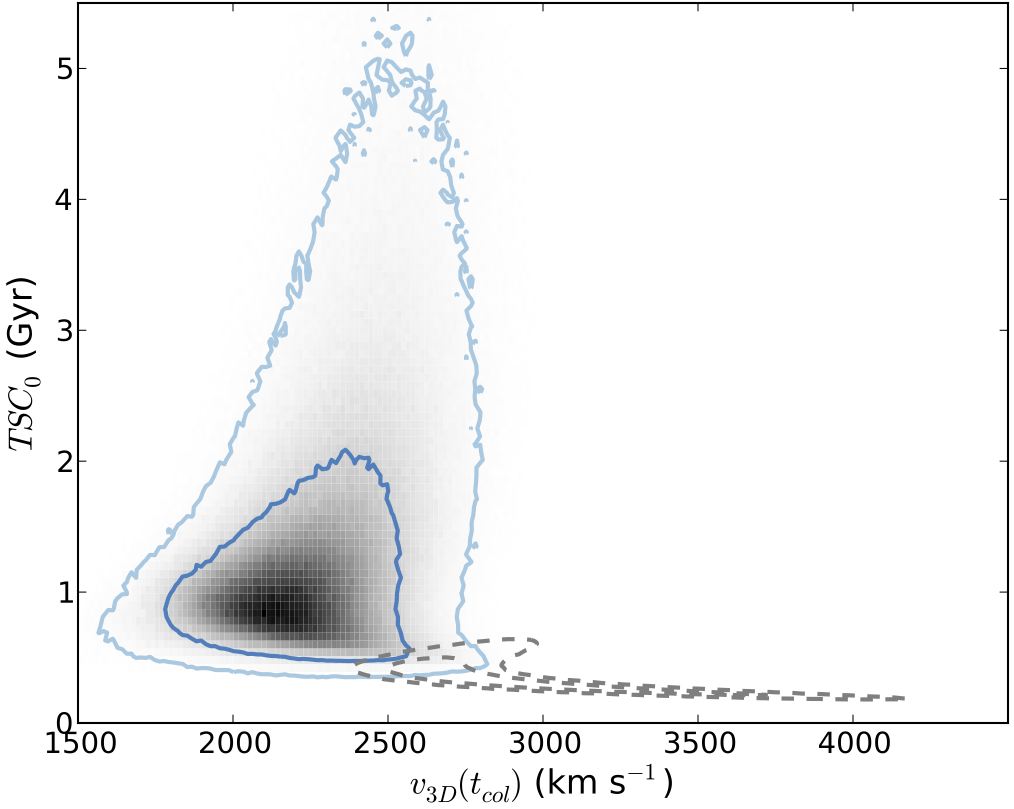}
\caption{
The posterior of the Musket Ball Cluster's $TSC_0$ and $v_{\rm 3D}(t_{\rm col})$ parameters is shown in grayscale with dark and light blue contours representing 68\% and 95\% confidence, respectively.
For comparison the gray dashed contours are the Bullet Cluster's 68\% and 95\% confidence intervals copied from Figure \ref{fig_bcTSC_added}.
The Musket Ball Cluster occupies a much different region of merger phase than the Bullet Cluster, having both a slower relative collision velocity and being observed in a much later stage of merger.
\label{fig_mbc_TSC}}
\end{figure}

\begin{figure}
\epsscale{1}
\plotone{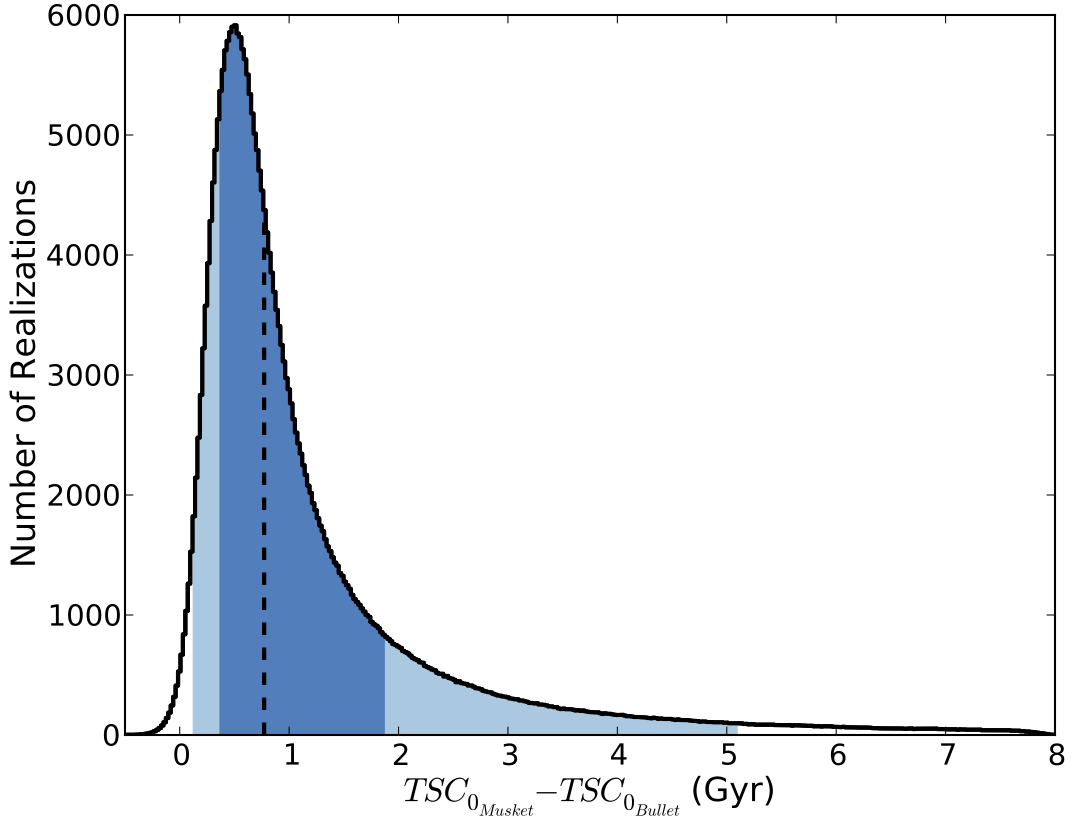}
\caption{
The histogram presents the $TSC_{\rm 0_{Musket}}-TSC_{\rm 0_{Bullet}}$ distribution from random draws of the respective marginalized $TSC_0$ distributions; showing that the Musket Ball Cluster merger is at least $0.8^{+1.2}_{-0.4}$\,Gyr ($3.4^{+3.8}_{-1.4}$ times) further progressed than the Bullet Cluster merger.
The black dashed line is the biweight-statistic location \citep{Beers:1982dp}, the 
dark and light blue regions denote the bias-corrected  68\% and 95\% lower and upper confidence limits, respectively.
\label{fig_TSC0comparison}}
\end{figure}

This more complete analysis confirms that the \object{Musket Ball Cluster} merger is both significantly slower and further progressed compared to the \object{Bullet Cluster}, see Figure \ref{fig_mbc_TSC}.
To estimate a lower limit on how much further progressed I perform an additional Monte Carlo analysis  for $TSC_{\rm 0_{Musket}}-TSC_{\rm 0_{Bullet}}$ assuming the marginalized $TSC_0$ distributions (see Appendices \ref{sec_bcresults} and \ref{sec_mbcresults}).
This is a lower limit since the \object{Musket Ball} observations, unlike the \object{Bullet Clusters} observations, cannot rule out the case that its subclusters have reached the apoapsis and are on a return trajectory (61\% of the realizations have $TSC_1$ less than the age of the Universe at $z=0.53$).
I find that  the \object{Musket Ball} is at least $0.8^{+1.2}_{-0.4}$\,Gyr ($3.4^{+3.8}_{-1.4}$ times) further progressed than the \object{Bullet Cluster}, see Figure \ref{fig_TSC0comparison}.
This is in line with the more approximate 3--5 times estimate of \citet{Dawson:2012dl}.
The \object{Musket Ball}'s relatively large $TSC_0$ means that it has potential for providing tighter constraints on $\sigma_{\rm DM}$, since the expected offset between the galaxies and dark matter will initially increase with increasing $TSC_0$.
However as noted in \S\ref{sec_intro}, given enough time the expected offset will decrease due the gravitational attraction between the galaxies and dark matter.
Also important in determining which cluster can provide the tightest $\sigma_{\rm DM}$ constraints is the fact the expected offset increases as a function of the cluster surface mass density and collision velocity, both of which are larger in the the case of the \object{Bullet Cluster} (compare Tables \ref{bulletresultparam} \& \ref{musketballresultparam}). 
Without running SIDM simulations it is difficult to know at what $TSC_0$ the offset reaches it maximum, or which merger parameters are most important for maximizing the offset.
The complete \object{Musket Ball Cluster} parameter estimates are summarized in Table \ref{musketballresultparam} and plotted in Appendix \ref{sec_mbcresults}.

\begin{deluxetable*}{lcccc}
\tablewidth{0pt}
\tablecaption{Musket Ball Cluster parameter estimates\label{musketballresultparam}}

\tablehead{
\colhead{Parameter}     & \colhead{Units}  & \colhead{Location\tablenotemark{a}}  & \colhead{68\% LCL--UCL\tablenotemark{b}} & \colhead{95\% LCL--UCL\tablenotemark{b}}
}
\startdata
M$_{200_1}$	&	$10^{14}$\,M$_\sun$	&	3.2	&	2.3	--	4.3	&	1.6	--	5.5	\\
M$_{200_2}$	&	$10^{14}$\,M$_\sun$	&	1.7	&	1.1	--	2.4	&	0.6	--	3.3	\\
$z_1$	&		&	0.5339	&	0.5333	--	0.5345	&	0.5326	--	0.5352	\\
$z_2$	&		&	0.5316	&	0.5305	--	0.5324	&	0.5294	--	0.5331	\\
$d_{\rm proj}$		&	Mpc	&	1.0	&	0.9	--	1.1	&	0.7	--	1.3	\\
$\alpha$	&	degree	&	48	&	28	--	67	&	13	--	78	\\
$d_{\rm 3D}$	&	Mpc	&	1.6	&	1.2	--	2.9	&	0.9	--	5.5	\\
$d_{\rm max}$	&	Mpc	&	2.1	&	1.5	--	3.8	&	1.1	--	7.3	\\
$v_{\rm 3D}(t_{\rm obs})$	&	km\,s$^{-1}$	&	670	&	390	--	1100	&	140	--	1500	\\
$v_{\rm 3D}(t_{\rm col})$ &	km\,s$^{-1}$	&	2300	&	2000	--	2500	&	1800	--	2800	\\
$TSC_0$	&	Gyr	&	1.1	&	0.7	--	2.4	&	0.5	--	5.8	\\
$TSC_1$\tablenotemark{c}	&	Gyr	&	3.5	&	2.0	--	7.2	&	1.4	--	12.0	\\
$T$	&	Gyr	&	4.8	&	2.9	--	10.4	&	2.2	--	22.7	\\
\enddata
\tablenotetext{a}{Biweight-statistic location \citep[see e.g.][]{Beers:1990kg}.}
\tablenotetext{b}{Bias-corrected lower and upper confidence limits, LCL and UCL respectively \citep[see e.g.][]{Beers:1990kg}.}
\tablenotetext{c}{61\% of the realizations with a valid $TSC_0$ (i.e.\,less than the age of the Universe at the cluster redshift) have a valid $TSC_1$, meaning that it is possible that the Musket Ball Cluster is being observed in the ``incoming'' state.}
\end{deluxetable*}

Note that just as a temporal prior was justified for the \object{Bullet Cluster} based on the observed shock front and increased temperature/mass estimate, I could apply a similar yet opposite prior to the \object{Musket Ball} since the temperature/mass estimate is consistent with the weak lensing inferred mass (additionally no shock front is observed).  
According to \citet{Randall:2002kk} if the cluster mass and inferred X-ray temperature or luminosity cluster mass are approximately the same then $TSC_0 \gtrsim 2 t_{\rm sc}$, which in the case of the \object{Musket Ball} means $TSC_0 \gtrsim 1.75$\,Gyr.
While this is consistent with my $TSC_0$ estimate for the \object{Musket Ball}, it is not entirely appropriate to apply this prior since the X-ray observations are relatively shallow and cannot confidently rule out a temperature and luminosity boost \citep{Dawson:2012dl}.
However it is conceivable that this line of reasoning would be applicable with deeper X-ray observations, either for the \object{Musket Ball} or similar dissociative mergers.

 
\section{Summary and Discussion}\label{summary}

I have introduced a new method for determining the dynamic properties and associated uncertainty of dissociative cluster mergers given only the most general merger observables: mass of each subcluster, redshift of each subcluster, and projected separation the subclusters.
I find that this method addresses the primary weaknesses of existing methods, namely enabling accurate parameter estimation and propagation of uncertainty near the collision state with a convergent solution achieved in $\sim$6\,CPU\,hours. 
I have confirmed that the two NFW halo model is capable of achieving the required 10\% level accuracy by direct comparison with an N-body hydrodynamic simulation.

In applying this method to the \object{Bullet Cluster} I not only determined its merger dynamic parameters but found that the bulk of uncertainty in these parameters is due to uncertainty in $\alpha$, the angle of the merger with respect to the plane of the sky.
Analyses that fail to account for the uncertainty in $\alpha$ (all existing N-body simulations of the \object{Bullet Cluster}) will significantly underestimate the uncertainty in their results.
This highlights the need to carefully select and model many possible realizations of the merger when trying to infer results from N-body simulations of a real merger (I discuss this further in \S\ref{sec_suggestions}).
I have also shown how ex post facto priors can easily be applied to the results of the default priors to further constrain the inferred dynamic properties.
In particular accurate measurement of the cluster gas properties can enable approximately a factor of two better constraint on the dynamic properties of the merger, principally through added constraint on the time scale of the merger.

I have also applied this method to the \object{Musket Ball Cluster}, validating the approximate results of \citet{Dawson:2012dl}.
Comparing the dynamic properties of the \object{Musket Ball} with those of the \object{Bullet} I have shown that the \object{Musket Ball} represents a significantly different volume of merger phase space.
The \object{Musket Ball Cluster}, being $3.4^{+3.8}_{-1.4}$ times further progressed than the \object{Bullet Cluster}, could potentially provide tighter constraints on $\sigma_{\rm DM}$ since the offset between galaxies and dark matter should initially increase with time post-merger for  $\sigma_{\rm DM}>0$.
And the larger the expected offset, the better the dark matter constraint when applying a method similar to \citet{Randall:2008hs}.

\subsection{Suggested Uses of Method}\label{sec_suggestions}

While a general method for determining the dynamics properties of merging clusters has numerous applications, several are worth noting.
As noted N-body simulations of specific merging clusters are computationally expensive; in particular one SIDM simulation of a single dissociative merger requires $\sim$1--10 million CPU\,hours (private communication, James Bullock). 
Thus it is currently unfeasible to simulate all confirmed dissociative mergers.
This method can be used to quickly determine which mergers provide the best $\sigma_{\rm DM}$ constraining power, enabling an efficient use of limited computational resources.

Additionally it is inappropriate to simply simulate one realization of a dissociative merger due to the broad range of merger phase space allowed by uncertainty in observed parameters, as discussed in detail in \S\ref{sec_defaultprior}.
Thus multiple simulations of each merger are required to properly represent the allowed phase space.
One could conceivably reduce the number of required simulations by using the results of this method to select representative merger realizations that uniformly sample the merger phase space of interest (e.g. cluster mass, $v_{\rm 3D}(t_{\rm col})$, and $TSC_0$);
then weight the results of each simulated realization by the integral of the corresponding local phase space PDF, as determined by this method.
For example, one could estimate the uncertainty distribution of the $\sigma_{\rm DM}$ constraint inferred from SIDM simulations of the \object{Bullet Cluster} by weighting the constraint from each realization, where a realization with  $v_{\rm 3D}(t_{\rm col})$=2800\,km\,s$^{-1}$ and $TSC_0$=0.4\,Gyr would receive greater weight than one with $v_{\rm 3D}(t_{\rm col})$=4000\,km\,s$^{-1}$ and $TSC_0$=0.2\,Gyr, see Figure \ref{fig_bcTSC_added}.
Thus the results of this method will not only inform efficient selection of realizations to model but will reduce the number of simulations required to properly sample the posterior PDF's.
Nevertheless SIDM simulations of mock clusters need to be performed to determine how much acceptable values of $\sigma_{\rm DM}$ affect the inferred merger dynamics properties.

General merger dynamic properties are also important for understanding  how cluster mergers relate to other physical phenomena, such as galaxy evolution and radio relics.  
It is well established that galaxy clusters play an important role in the evolution of their member galaxies, but it is still unclear whether cluster mergers trigger star formation \citep[e.g.][]{Miller:2003kx,Owen:2005dx,Ferrari:2005es,Hwang:2009ip}, quench it \citep{Poggianti:2004ca}, or have no immediate effect \citep{Chung:2010ds}.
Studying mergers at different $TSC$ may resolve these seemingly conflicting results by  discriminating between slow-working processes (e.g.\,galaxy harassment or strangulation) and fast-acting process (e.g.\,ram pressure stripping). 
Similarly, studying global merger dynamic properties may resolve the mystery of why many mergers have associated radio relics \citep[e.g.][]{Barrena:2009to, vanWeeren:2011ko} yet others don't \citep[e.g.][]{Russell:2011hn}.

\subsection{Extensions to the Method}\label{sec_extensions}

While this method has advantages over existing methods there is room for considerable improvement.
For example the method could be improved through the elimination of some of the simplifying assumptions of the model (see \S\ref{sec_model}).
One could attempt to incorporate subcluster mass accretion physics in a manner similar to the work of \citet{Angus:2007em} or attempt to account for the possibility of a non-zero impact parameter.
To incorporate the latter one must: 1) add angular momentum terms to the equations of motion,  which is entirely feasible, and 2) prescribe a reasonable impact parameter prior.
\citet{Randall:2002kk} nicely outline how to determine an impact parameter PDF for halo mergers of variable mass by utilizing the PDF of the dimensionless \emph{spin parameter}, determined from linear theory of the growth of structure \citep{Peebles:1993vp} and simulations \citep{Bullock:2001kb}.
However, this prior should be adjusted to account for the samount of gas dissociated during the observed merger, since this amount will decrease as the impact parameter increases.
Without a systematic study of various mergers in hydrodynamic simulations it is unclear exactly what adjustment an observed large dissociation of gas should infer.

Another significant extension to the model could be the inclusion of SIDM physics.  
As mentioned in the previous section, one of the promising uses of this method is to suggest which mergers might provide the best $\sigma_{\rm DM}$ constraining power.  
However one could take this a step further by including an analytic treatment of SIDM physics \citep[e.g.\,][]{Markevitch:2004dl}, thereby enabling analytic estimates of $\sigma_{\rm DM}$ relevant effects for a given merger.
Then this method could be used in conjunction with observed dissociative mergers to place direct constraints on $\sigma_{\rm DM}$.
Due to the increased complexity of the physics involved
it would be necessary to verify this extension with SIDM N-body simulations.


\emph{Note:} W. Dawson has made Python code implementing the discussed Monte Carlo method openly available at git://github.com/MCTwo/MCMAC.git.
He has also made all supporting work to this paper openly available at 
git://github.com/wadawson/merging-cluster-dynamics-paper.git.

\acknowledgments

I thank my adviser David Wittman who has always encouraged my research-freewill, while at the same time providing invaluable input and correcting guidance.
Our many fruitful discussions have touched every aspect of this work.
I also thank the Deep Lens Survey --- in particular Perry Gee --- for access to the 2007 Keck LRIS spectra, as well as Perry Gee and Brian Lemaux for assistance in reduction of the 2011 Keck DEIMOS spectra.  
I am grateful to  Jack P.\,Hughes for discussions on constraining the \object{Bullet Cluster}'s TSC by the observed X-ray shock feature and boosted temperature/luminosity, and
Reinout J.\,van Weeren for discussions on the possibilities of constraining the dynamics and geometry of mergers using radio relics.
Finally I would like to thank Maru{\v s}a Brada{\v c}, Ami Choi, James Jee, Phil Marshall, Annika Peter, Michael Schneider, Reinout van Weeren, and David Wittman for their substantial reviews of earlier drafts.

Support for this work was provided by NASA through Chandra Award Number GO1-12171X issued by CXO Center, which is operated by the SAO for and on behalf of NASA under contract NAS8-03060.  Support for program number GO-12377 was provided by NASA through a grant from STScI, which is operated by the Association of Universities for Research in Astronomy, Inc., under NASA contract NAS5-26555.
Support for this work was also provide by Graduate Research Fellowships through the University of California, Davis.
I would also like to acknowledge the sultry universe who hides her secrets well.


{\it Facilities:} \facility{CXO (ACIS-I)}, \facility{HST (ACS)}, \facility{Keck:I (LRIS)}, \facility{Keck:2 (DEIMOS)}, \facility{Mayall (MOSAIC 1 \& 1.1)}, \facility{Subaru (Suprime-Cam)}, \facility{SZA}.


\clearpage

\appendix
%
%
%
%
%
\section{Potential Energy of Two Truncated NFW Halos}\label{potentialsec}

Generically the potential energy of a two-halo system with center to center separation $r$ is 
\begin{equation}
V(r) = \int \Phi_1(r') dm_2,
\label{potint}
\end{equation}
where $\Phi_1(r')$ is the gravitational potential of halo 1 as a function of radial distance $r'$ from the center of the halo 1 to the mass element of halo 2, $dm_2$.  
I derive $\Phi_1(r)$ for the case of a truncated NFW halo in \S \ref{nfwpotential}.
The integral of equation \ref{potint} can be approximated as a summation over $N \times N$ mass elements, $m_{2_{ij}}$, each with area $dr\times d\theta$, where $i$ and $j$ range from $0 \rightarrow N-1$,
\begin{displaymath}
V(r) \approx \sum_{i=0}^{N-1}\sum_{j=0}^{N-1} \Phi_1(r'_{ij}+\epsilon) m_{2_{ij}},
\end{displaymath}
where $r'_{ij}$ is the distance from the center of halo 1 to the 2$^{\rm nd}$ halo's mass element $m_{2_{ij}}$, as derived in \S\ref{masselements}, and $\epsilon$ is the softening length which reduces the effects of artificial singularities.

\subsection{Truncated NFW Gravitational Potential}\label{nfwpotential}

For an axially symmetric mass distribution the potential can be expressed as a series of Legendre Polynomials

\begin{equation}
\Phi_n(r) = -\frac{2\pi G}{(n+1/2)r^{n+1}} \int_{0}^{r} r'^{n+2} \rho_{n}(r')\,dr' -\frac{2\pi G r^n}{n+1/2} \int_{r}^{\infty} r'^{1-n} \rho_n(r')\,dr' 
\label{axsympot}
\end{equation}
where
\begin{equation}
\rho_n(r) = (n+1/2) \int_{0}^{\pi} \rho(r,\theta) P_n(\cos \theta) \sin \theta \,d\theta.
\label{lengden}
\end{equation}

Assuming a spherical NFW halo
\begin{displaymath}
\rho_{\rm NFW}(r) = \frac{\rho_s}{r/r_s(1+r/r_s)^2}
\end{displaymath}
only the zero$^{\rm th}$ order term of Equation \ref{lengden} remains
\begin{displaymath}
\rho_{\rm NFW}(r) = \rho_0(r)
\end{displaymath}
and Equation \ref{axsympot} reduces to
\begin{eqnarray}
\Phi_{\rm NFW}(r) & = & -\frac{4\pi G}{r}\int_{0}^{r} r'^2 \rho_{\rm NFW}(r')\,dr' - 4\pi G \int_{r}^{\infty}r' \rho_{\rm NFW}(r')\,dr' \nonumber\\
\Phi_{\rm NFW}(r) & = & -\frac{4\pi G \rho_s}{r} \left[ \int_{0}^{r} \frac{r'^2}{r'/r_s(1+r'/r_s)^2}\,dr' + r \int_{r}^{\infty} \frac{r'}{r'/r_s(1+r'/r_s)^2}\,dr' \right]. \nonumber
\end{eqnarray}
Since I truncate the NFW halo at $r_{200}$ the $\infty$ in the second integral becomes $r_{200}$ and
\begin{equation}
	\Phi_{\rm NFW_T}(r)=
	\begin{cases}
 		-\frac{4\pi G}{r} \rho_s r_s^3 \left[ \ln (1+r/r_s)-\frac{r}{r_s+r_{200}} \right], &\text{if $r\leq r_{200}$;}\\
  		-\frac{G M_{200}}{r}, &\text{if $r>r_{200}$.}
  	\end{cases}
\end{equation}

\subsection{Mass Elements of a Truncated NFW Halo}\label{masselements}

Given the differential mass elements for a spherically symmetric halo
\begin{displaymath}
dm = 2\pi\rho(r,\theta) r^2 \sin(\theta) d\theta\,dr,
\end{displaymath}
and discretizing the mass into elements with lengths $\delta r = r_{200_2}/N$ and $\delta\theta = \pi/N$ the halo 2 mass elements are given by
\begin{displaymath}
m_{ij} = 2\pi \int_{i\,\delta r}^{(i+1)\delta r} \int_{j\,\delta\theta}^{(j+1)\delta\theta} \rho(r') r'^2 \sin(\theta') d\theta'\,dr'.
\end{displaymath}
For an NFW halo this becomes
\begin{displaymath}
m_{ij} = 2\pi\rho_s r_s^3 \left[\cos(j\,\delta\theta)-\cos\left((j+1)\delta\theta\right) \right] \left[ \left( 1+\frac{(i+1)\delta r}{r_s}\right)^{-1} - \left(1+\frac{i\,\delta r}{r_s}\right)^{-1} + \ln\left[\frac{(i+1)\delta r+r_s}{i\,\delta r+r_s}\right]\right].
\end{displaymath}

\clearpage

\section{Bullet Cluster Result Plots}\label{sec_bcresults}

\begin{figure}[b]
\epsscale{1}
\plotone{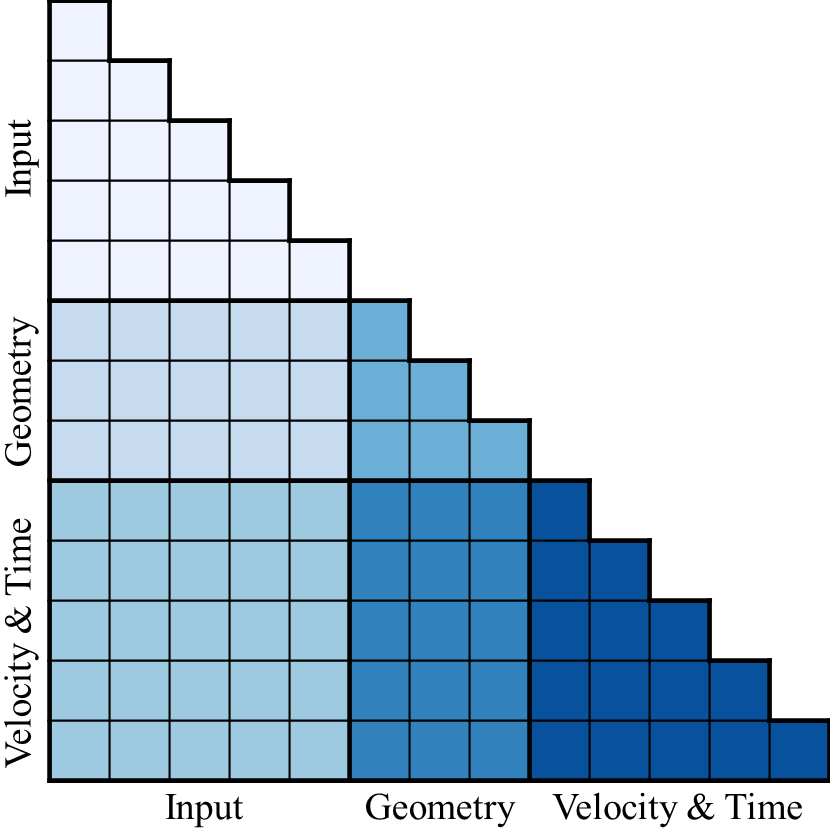}
\caption{
For ease of display the results array is divided into six subplots, Figures \ref{fig_bc_inin}--\ref{fig_bc_geovt}.
The Input parameters consist of: M$_{200_1}$, M$_{200_2}$, $z_1$, $z_2$,	and $d_{\rm proj}$.
The calculated Geometry parameters consist of: $\alpha$, $d_{\rm 3D}$, and $d_{\rm max}$.
The calculated Velocity \& Time parameters consist of:  $v_{\rm 3D}(t_{\rm obs})$, $v_{\rm 3D}(t_{\rm col})$, $TSC_0$, $TSC_1$, and $T$.
\label{fig_resultsarray}}
\end{figure}

This section contains the parameter results array plots for the Bullet Cluster case including the added temporal prior of \S\ref{sec_addedprior}.
For ease of display the parameters are grouped in three categories (\emph{Input}, \emph{Geometry}, and \emph{Velocity \& Time}) resulting in a six subplot results array, see Figure \ref{fig_resultsarray}.
The \emph{Input} parameters consist of: M$_{200_1}$, M$_{200_2}$, $z_1$, $z_2$,	and $d_{\rm proj}$, where halo 1 refers to the ``main'' subcluster and halo 2 refers to the ``bullet'' subcluster.
The \emph{Geometry} parameters consist of the randomly drawn $\alpha$, and calculated $d_{\rm 3D}$, and $d_{\rm max}$.
The calculated \emph{Velocity \& Time} parameters consist of:  $TSC_0$, $TSC_1$, and $T$.

\clearpage

\begin{figure}
\epsscale{1}
\plotone{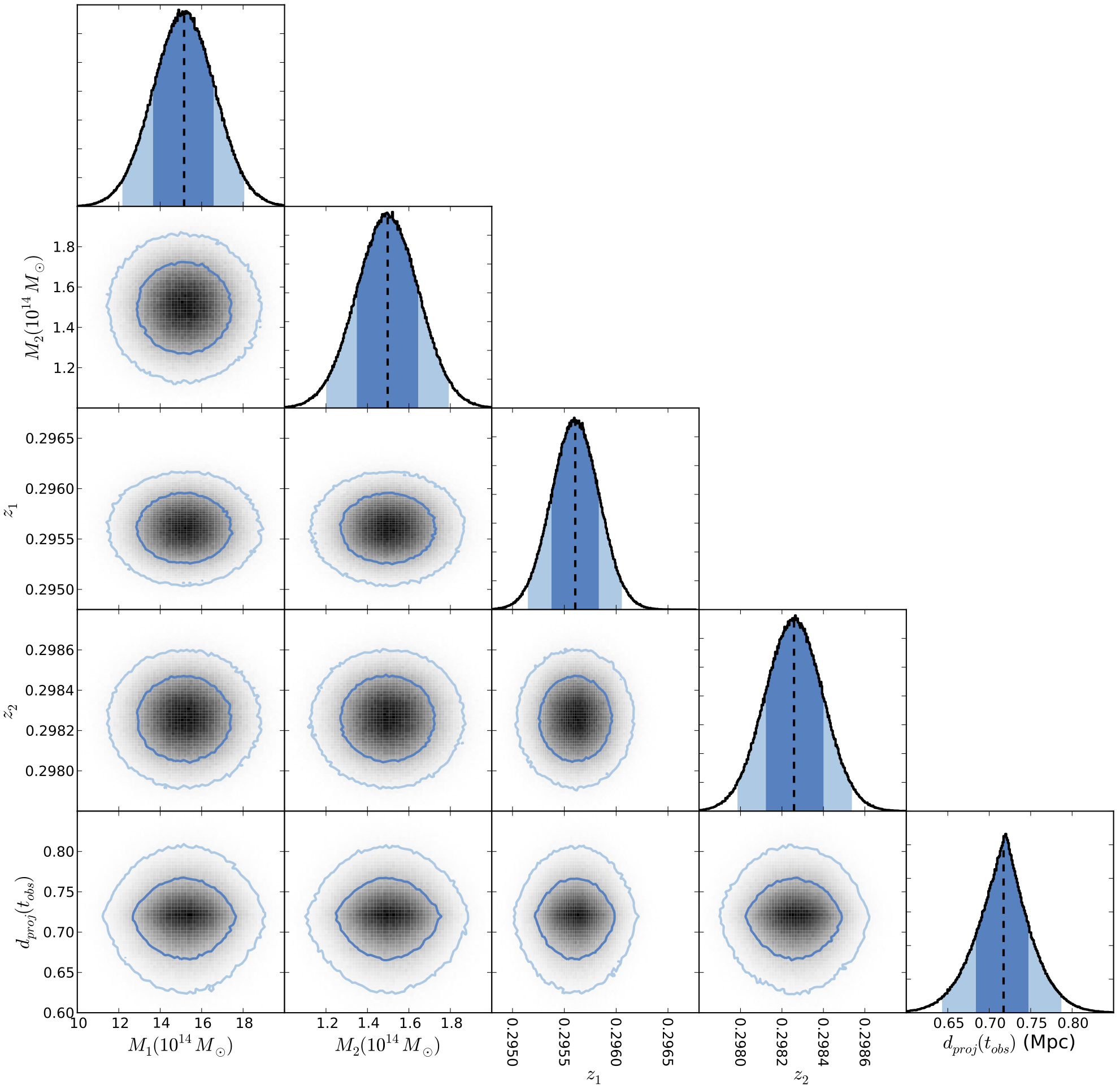}
\caption{Bullet Cluster marginalized \emph{Input vs.\,Input} parameters result plots, for the case including the added temporal prior of \S\ref{sec_addedprior}. Dark and light blue colors correspond to 68\% and 95\% confidence intervals, respectively.  The black dashed line is the biweight-statistic location \citep{Beers:1982dp}. 
\label{fig_bc_inin}}
\end{figure}

\begin{figure}
\epsscale{1}
\plotone{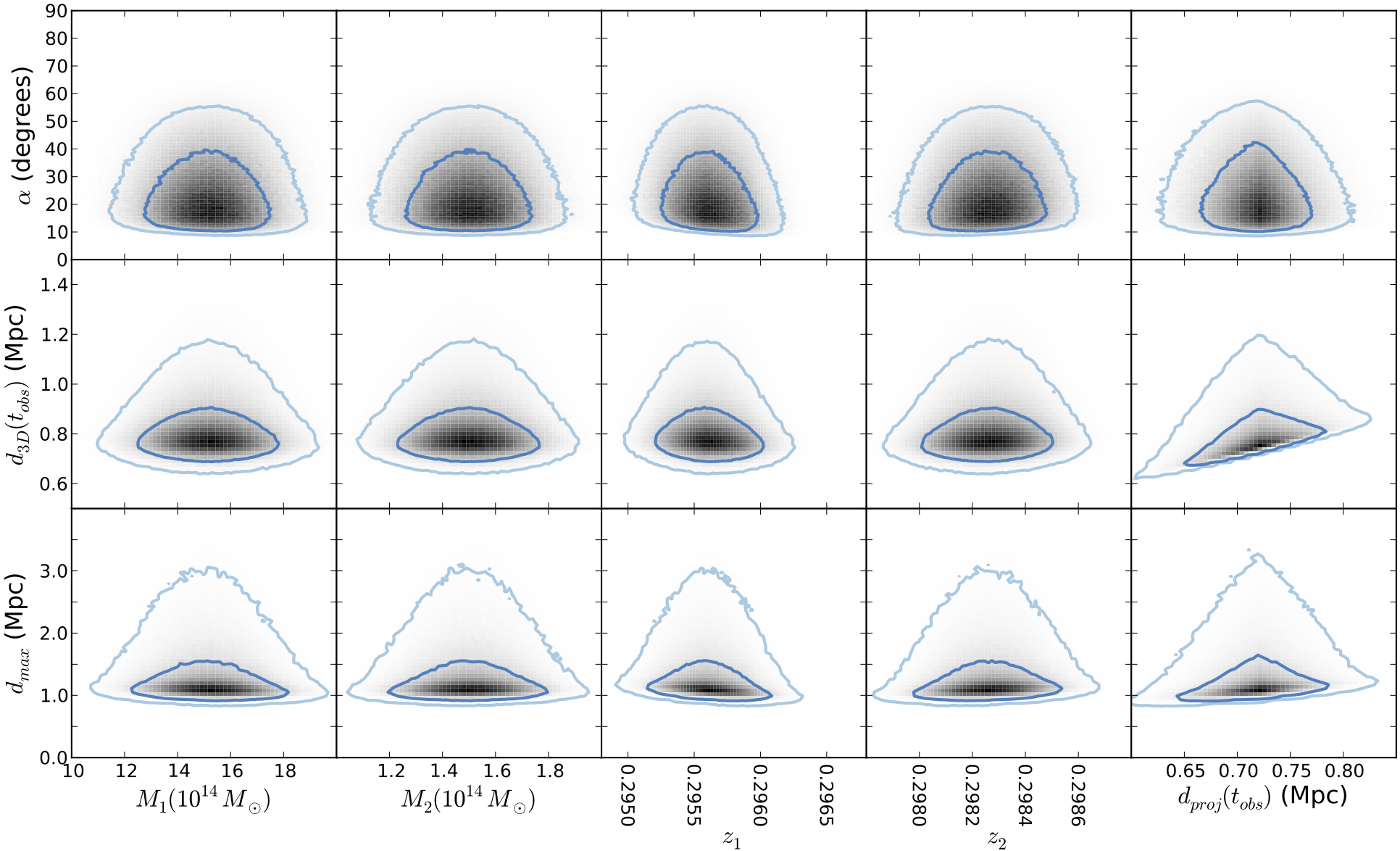}
\caption{Bullet Cluster marginalized \emph{Input vs.\,Geometry} parameters result plots, for the case including the added temporal prior of \S\ref{sec_addedprior}.  Dark and light blue colors correspond to 68\% and 95\% confidence intervals, respectively.
\label{fig_bc_ingeo}}
\end{figure}

\begin{figure}
\epsscale{1}
\plotone{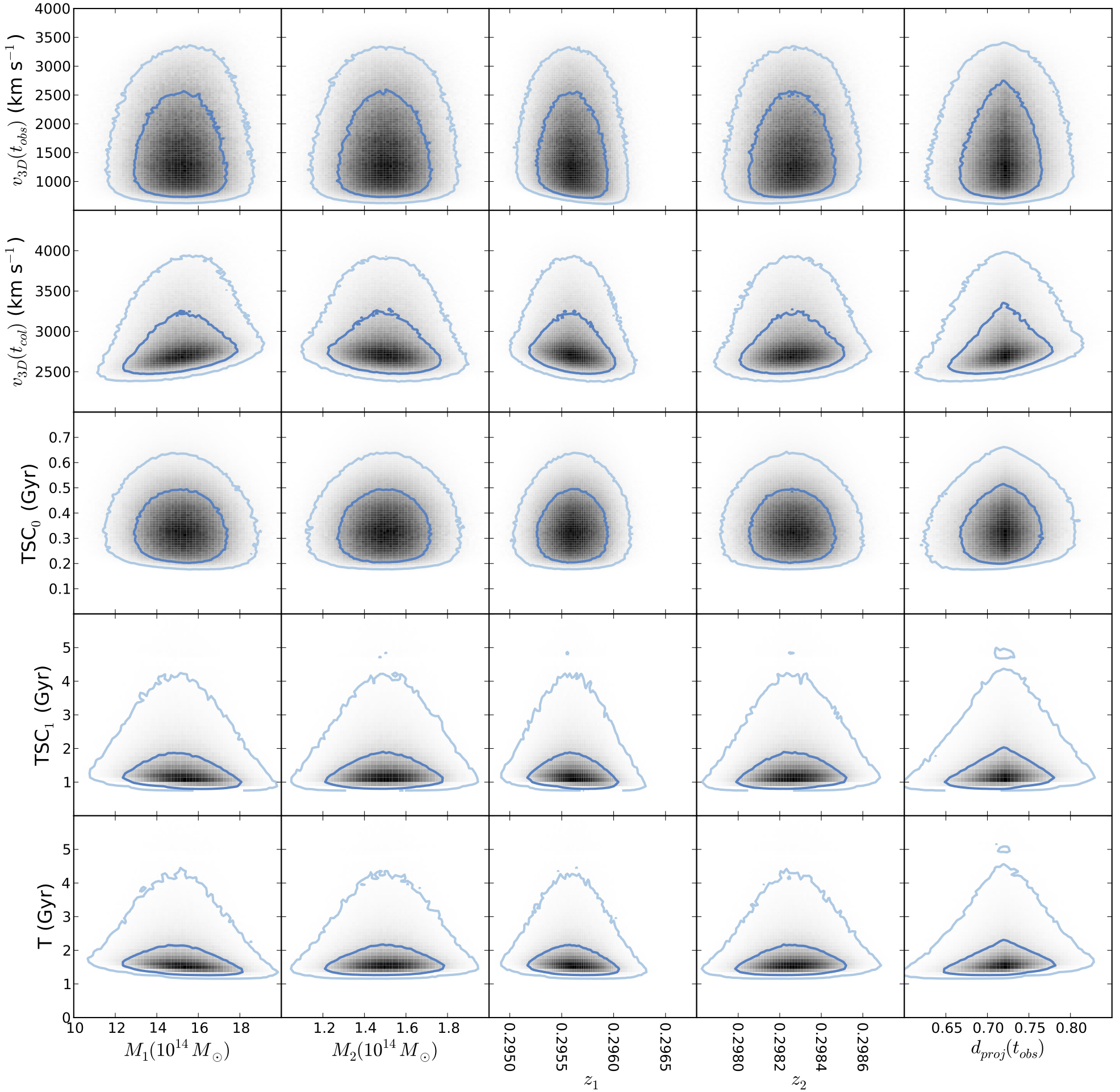}
\caption{Bullet Cluster marginalized \emph{Input vs.\,Velocity \& Time} parameters result plots, for the case including the added temporal prior of \S\ref{sec_addedprior}.  Dark and light blue colors correspond to 68\% and 95\% confidence intervals, respectively.
\label{fig_bc_invt}}
\end{figure}

\begin{figure}
\epsscale{1}
\plotone{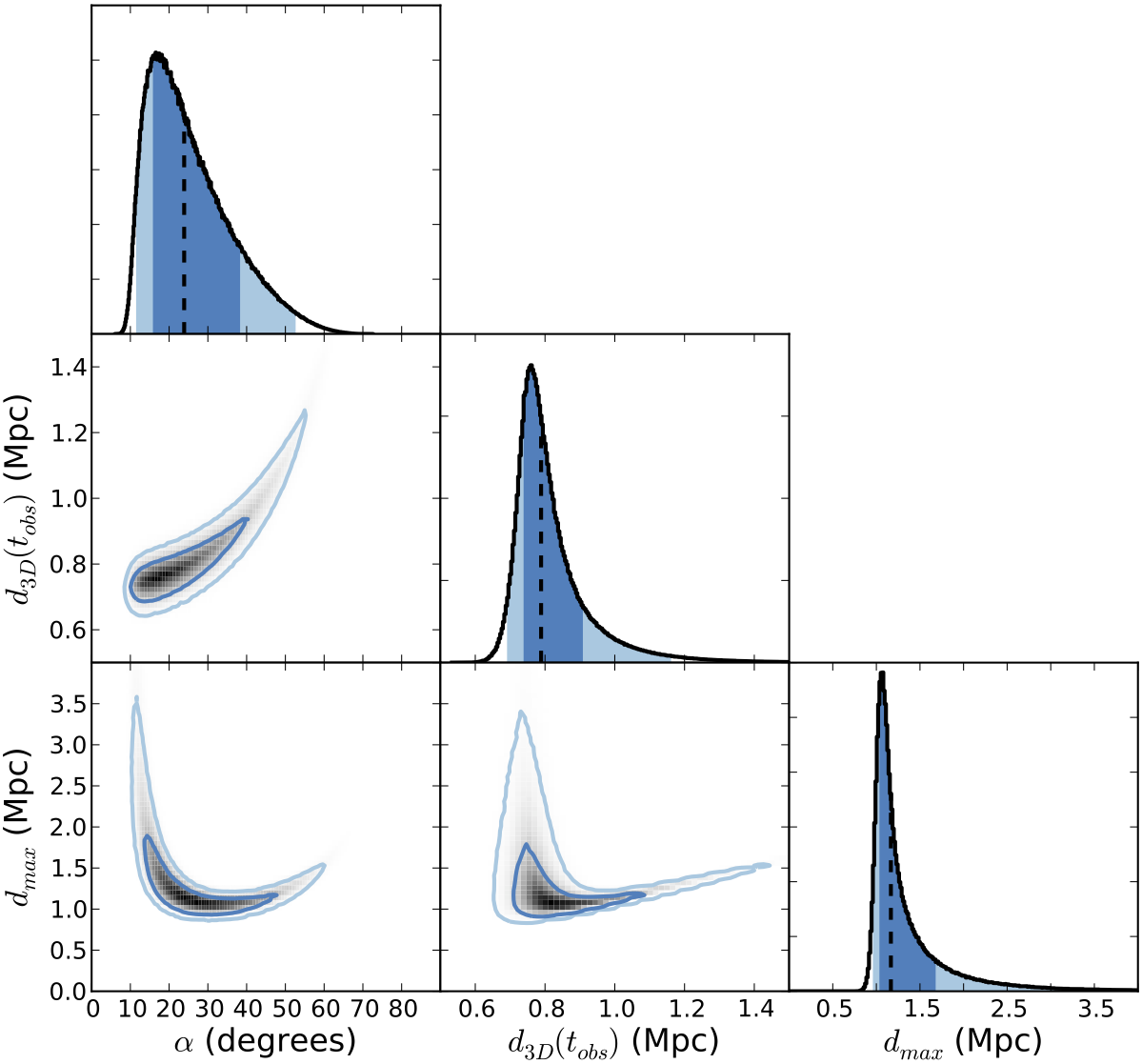}
\caption{Bullet Cluster marginalized \emph{Geometry vs.\,Geometry} parameters result plots, for the case including the added temporal prior of \S\ref{sec_addedprior}.  Dark and light blue colors correspond to 68\% and 95\% confidence intervals, respectively.  The black dashed line is the biweight-statistic location \citep{Beers:1982dp}.
\label{fig_bc_geogeo}}
\end{figure}

\begin{figure}
\epsscale{1}
\plotone{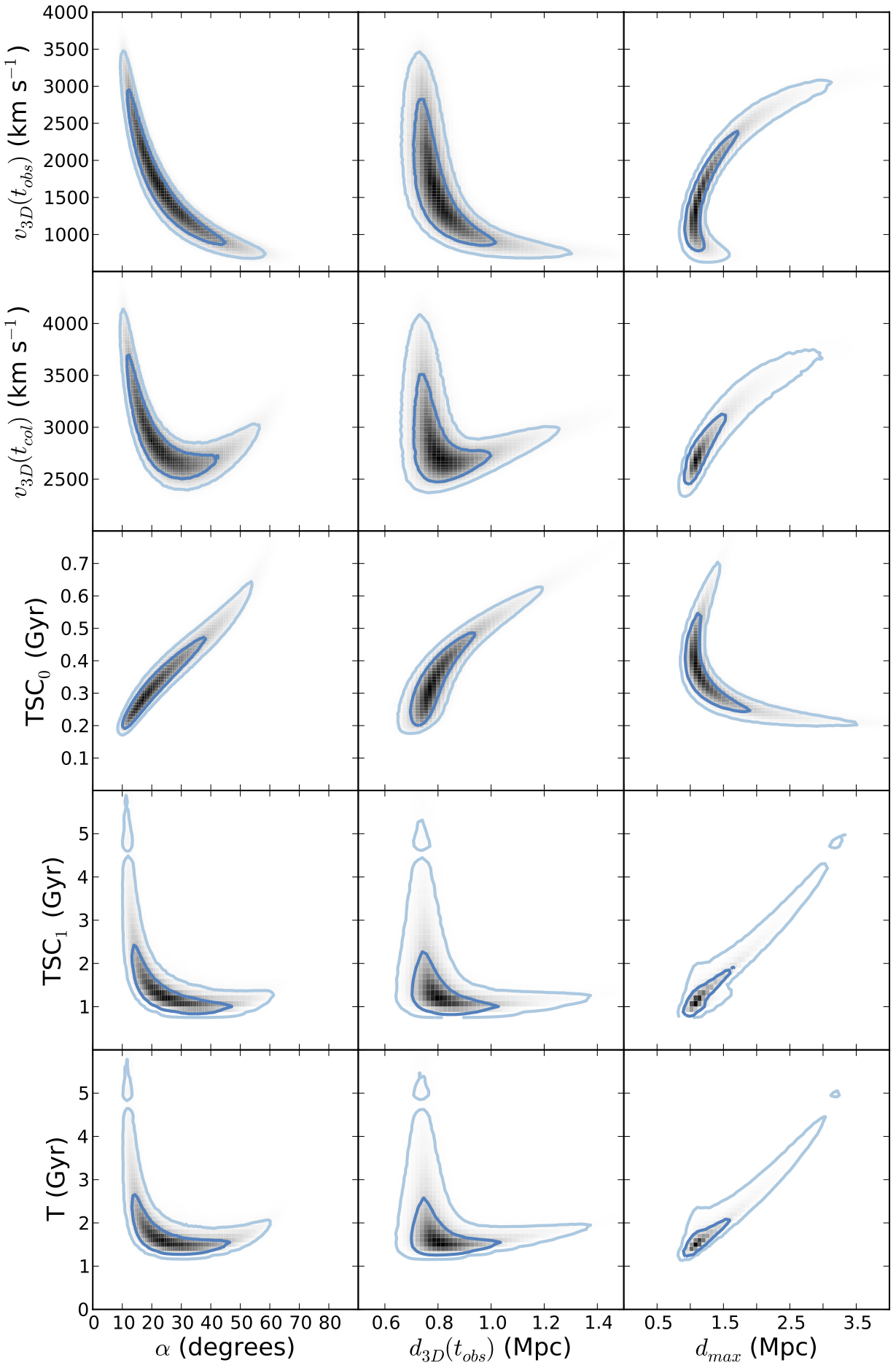}
\caption{Bullet Cluster marginalized \emph{Geometry vs.\,Velocity \& Time} parameters result plots, for the case including the added temporal prior of \S\ref{sec_addedprior}.  Dark and light blue colors correspond to 68\% and 95\% confidence intervals, respectively.
\label{fig_bc_geovt}}
\end{figure}

\begin{figure}
\epsscale{1}
\plotone{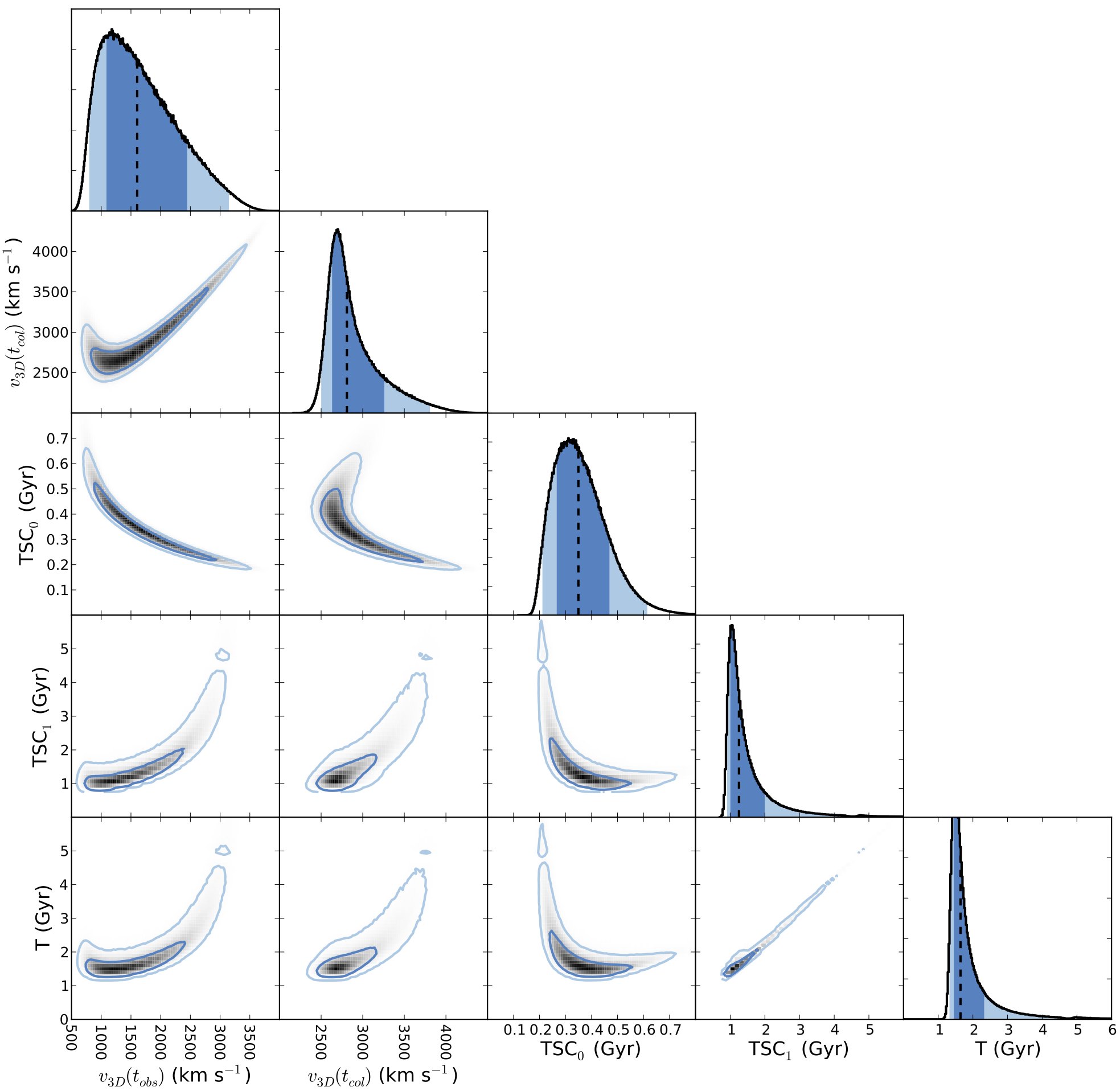}
\caption{Bullet Cluster marginalized \emph{Velocity \& Time vs.\,Velocity \& Time} parameters result plots, for the case including the added temporal prior of \S\ref{sec_addedprior}.  Dark and light blue colors correspond to 68\% and 95\% confidence intervals, respectively.  The black dashed line is the biweight-statistic location \citep{Beers:1982dp}.
\label{fig_bc_vtvt}}
\end{figure}
\clearpage

\section{Musket Ball Cluster Result Plots}\label{sec_mbcresults}

This section contains the parameter results array plots for the \object{Musket Ball Cluster}.
Similar to \S\ref{sec_bcresults} the parameters are grouped in three categories (\emph{Input}, \emph{Geometry}, and \emph{Velocity \& Time}) resulting in a six subplot results array, see Figure \ref{fig_resultsarray}.
The \emph{Input} parameters consist of: M$_{200_1}$, M$_{200_2}$, $z_1$, $z_2$,	and $d_{\rm proj}$, where halo 1 refers to the ``south'' subcluster and halo 2 refers to the ``north'' subcluster.
The calculated \emph{Geometry} parameters consist of: $\alpha$, $d_{\rm 3D}$, and $d_{\rm max}$.
The calculated Velocity \& Time parameters consist of:  $v_{\rm 3D}(t_{\rm obs})$, $v_{\rm 3D}(t_{\rm col})$, $TSC_0$, $TSC_1$, and $T$.

\begin{figure}[b]
\epsscale{1}
\plotone{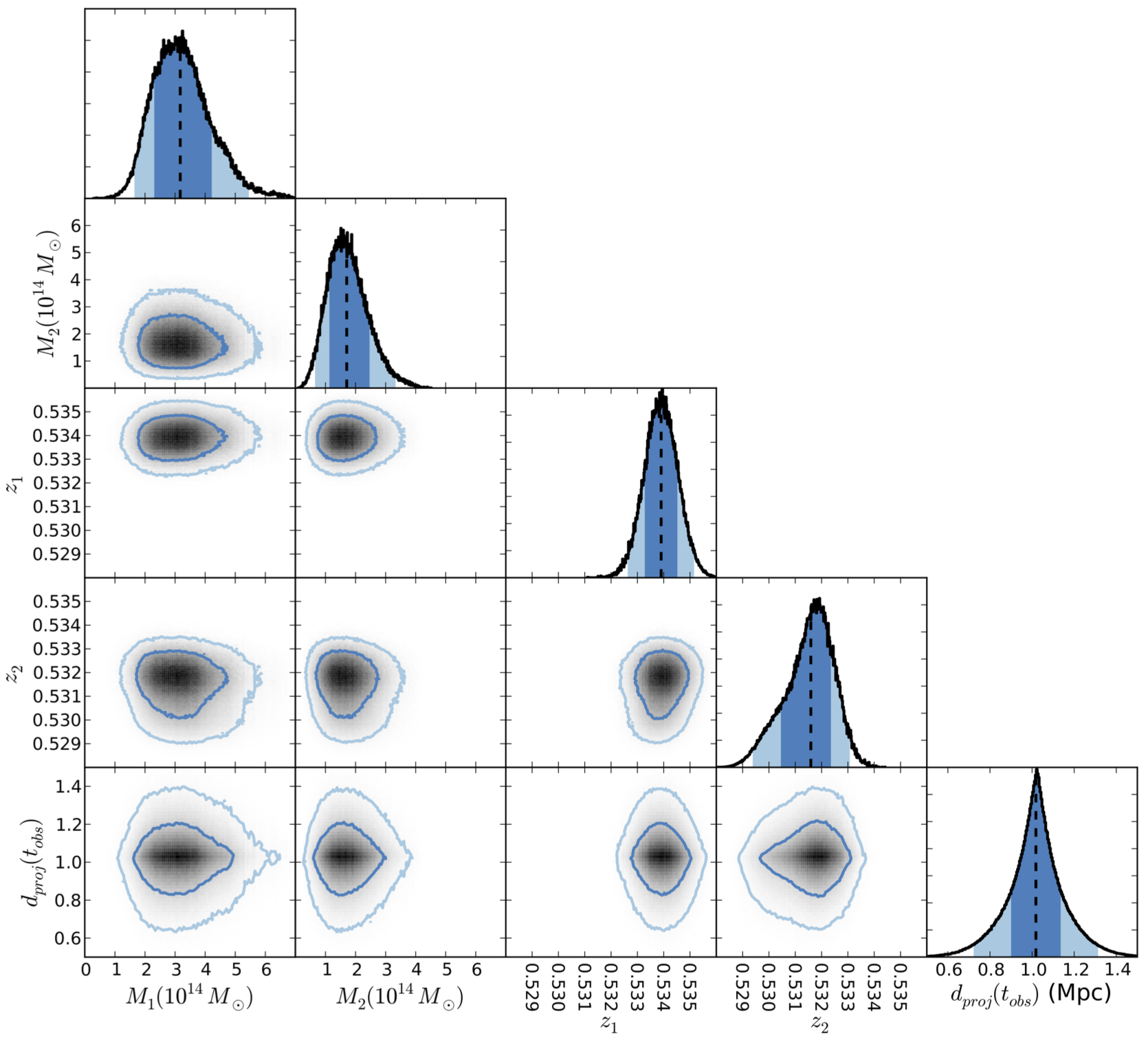}
\caption{Musket Ball Cluster marginalized \emph{Input vs.\,Input} parameters result plots.  Dark and light blue colors correspond to 68\% and 95\% confidence intervals, respectively.  The black dashed line is the biweight-statistic location \citep{Beers:1982dp}.
\label{musket_inin}}
\end{figure}

\begin{figure}
\epsscale{1}
\plotone{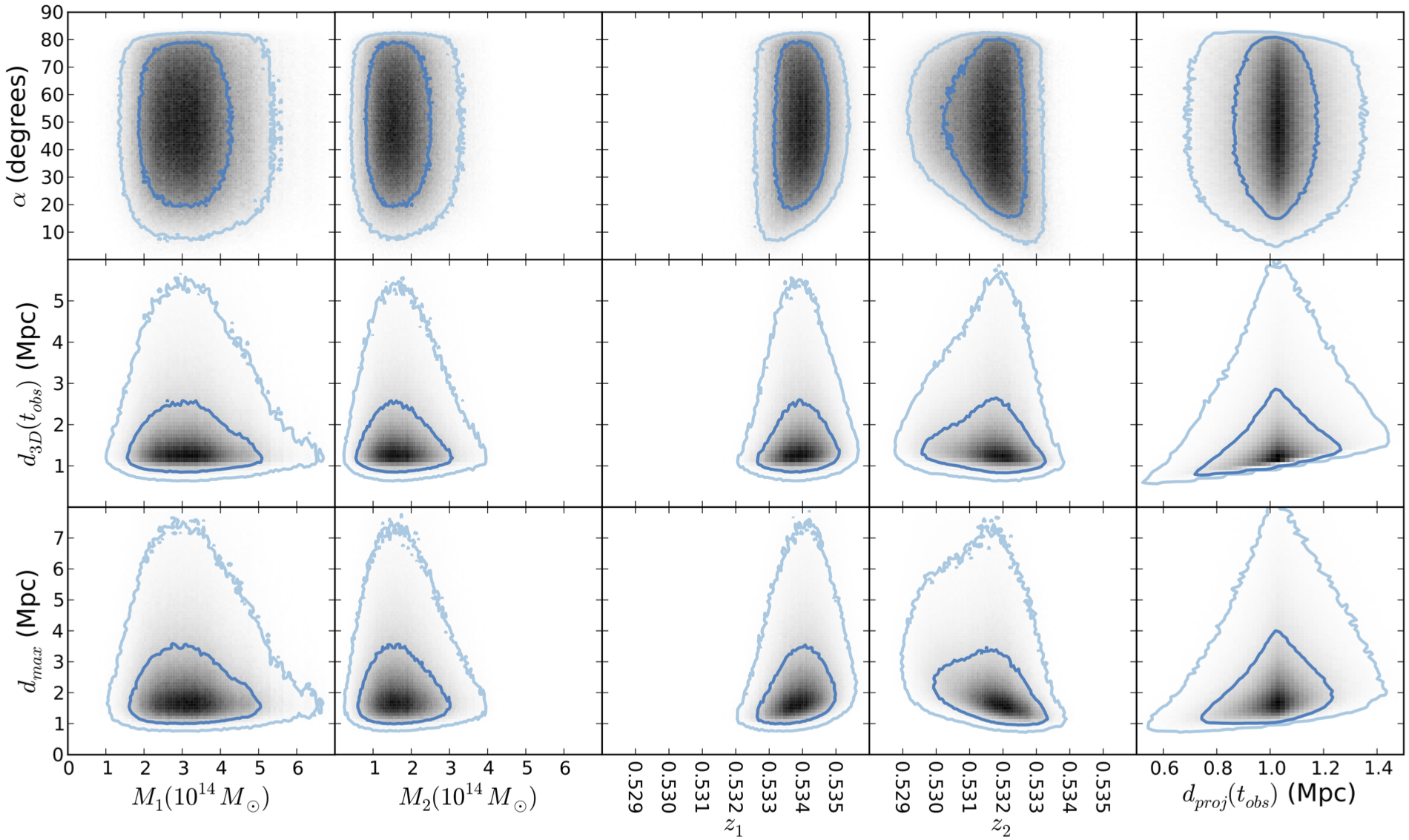}
\caption{Musket Ball Cluster marginalized \emph{Input vs.\,Geometry} parameters result plots.  Dark and light blue colors correspond to 68\% and 95\% confidence intervals, respectively.
\label{musket_ingeo}}
\end{figure}

\begin{figure}
\epsscale{1}
\plotone{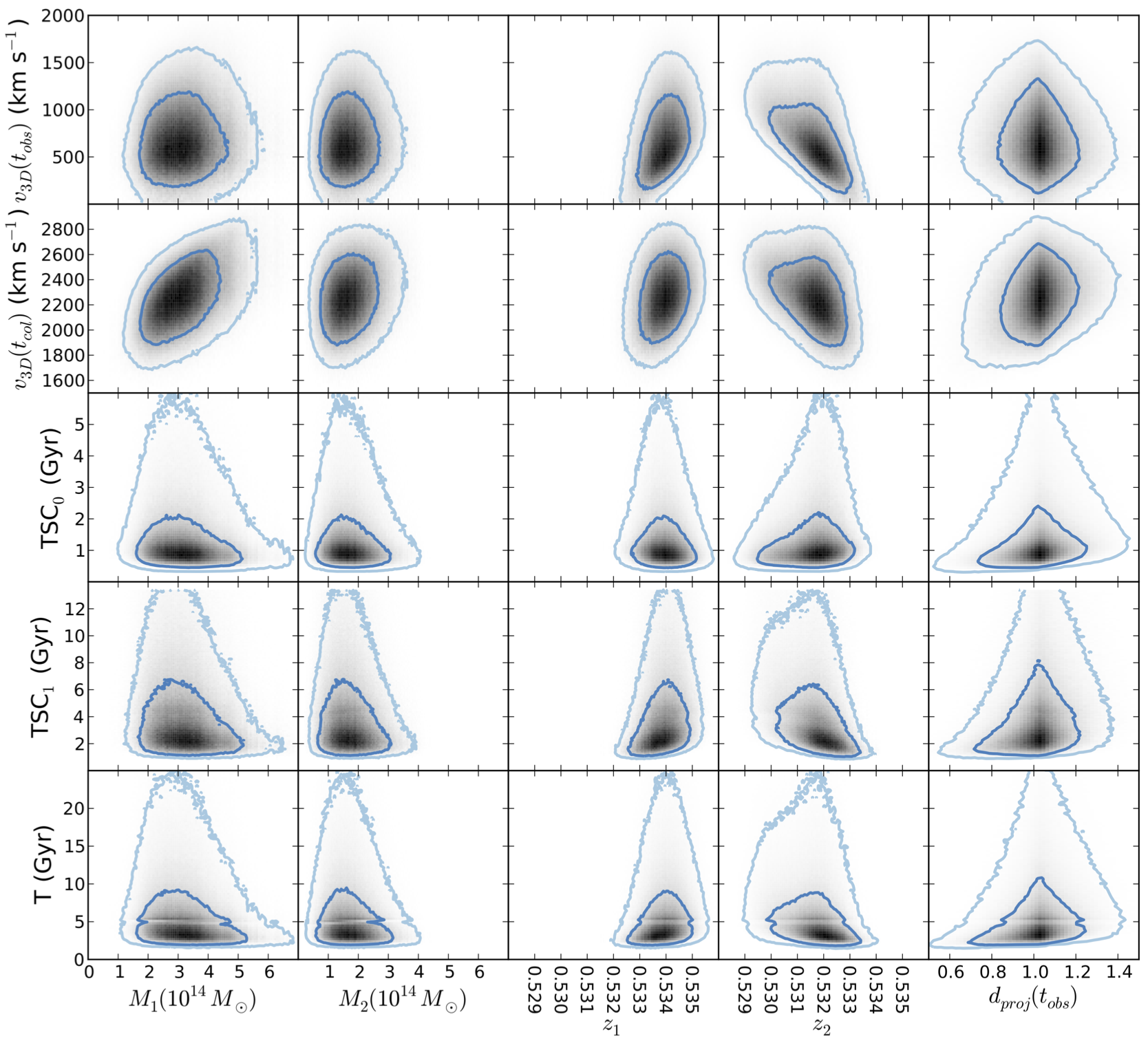}
\caption{Musket Ball Cluster marginalized \emph{Input vs.\,Velocity \& Time} parameters result plots.  Dark and light blue colors correspond to 68\% and 95\% confidence intervals, respectively.
\label{musket_invt}}
\end{figure}

\begin{figure}
\epsscale{1}
\plotone{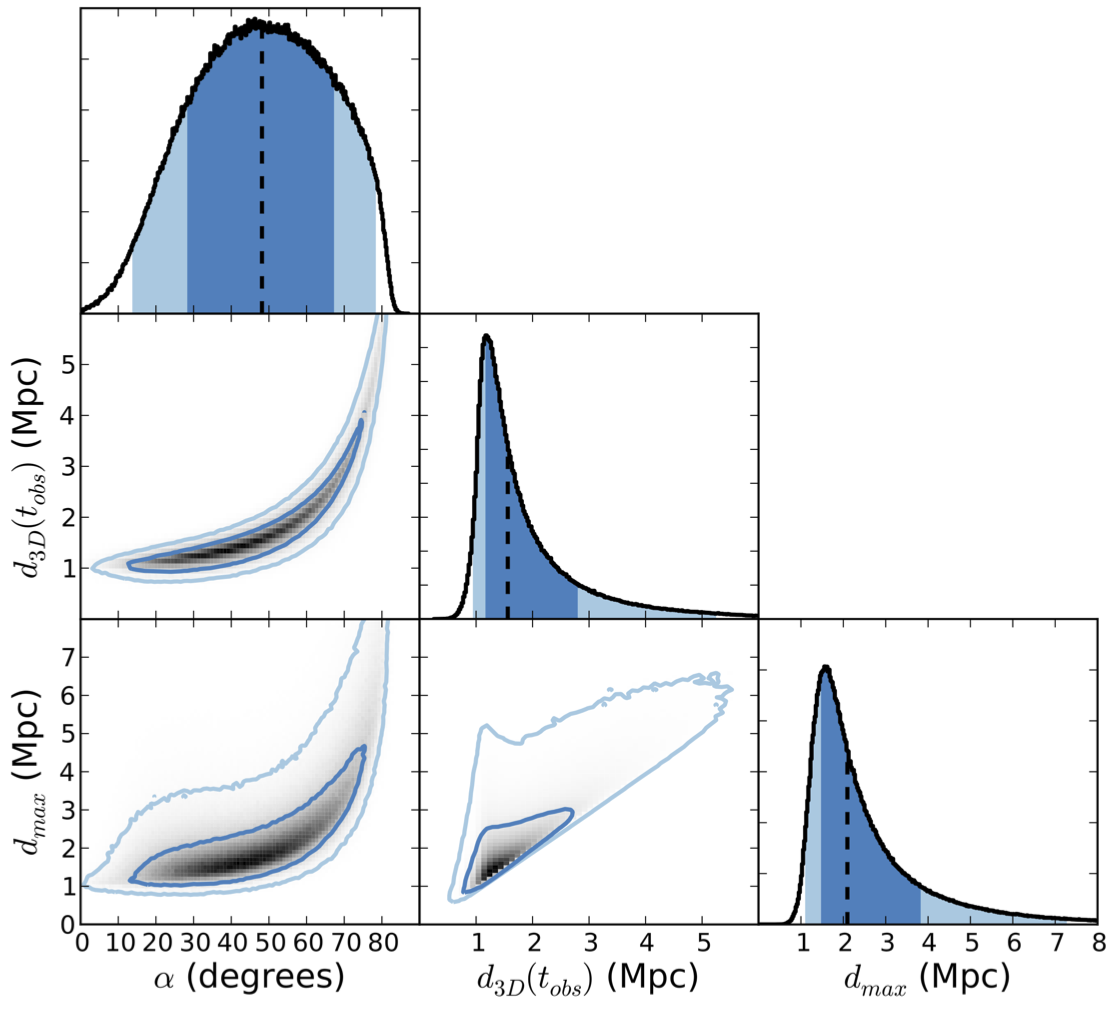}
\caption{Musket Ball Cluster marginalized \emph{Geometry vs.\,Geometry} parameters result plots.  Dark and light blue colors correspond to 68\% and 95\% confidence intervals, respectively.  The black dashed line is the biweight-statistic location \citep{Beers:1982dp}.
\label{musket_geogeo}}
\end{figure}

\begin{figure}
\epsscale{1}
\plotone{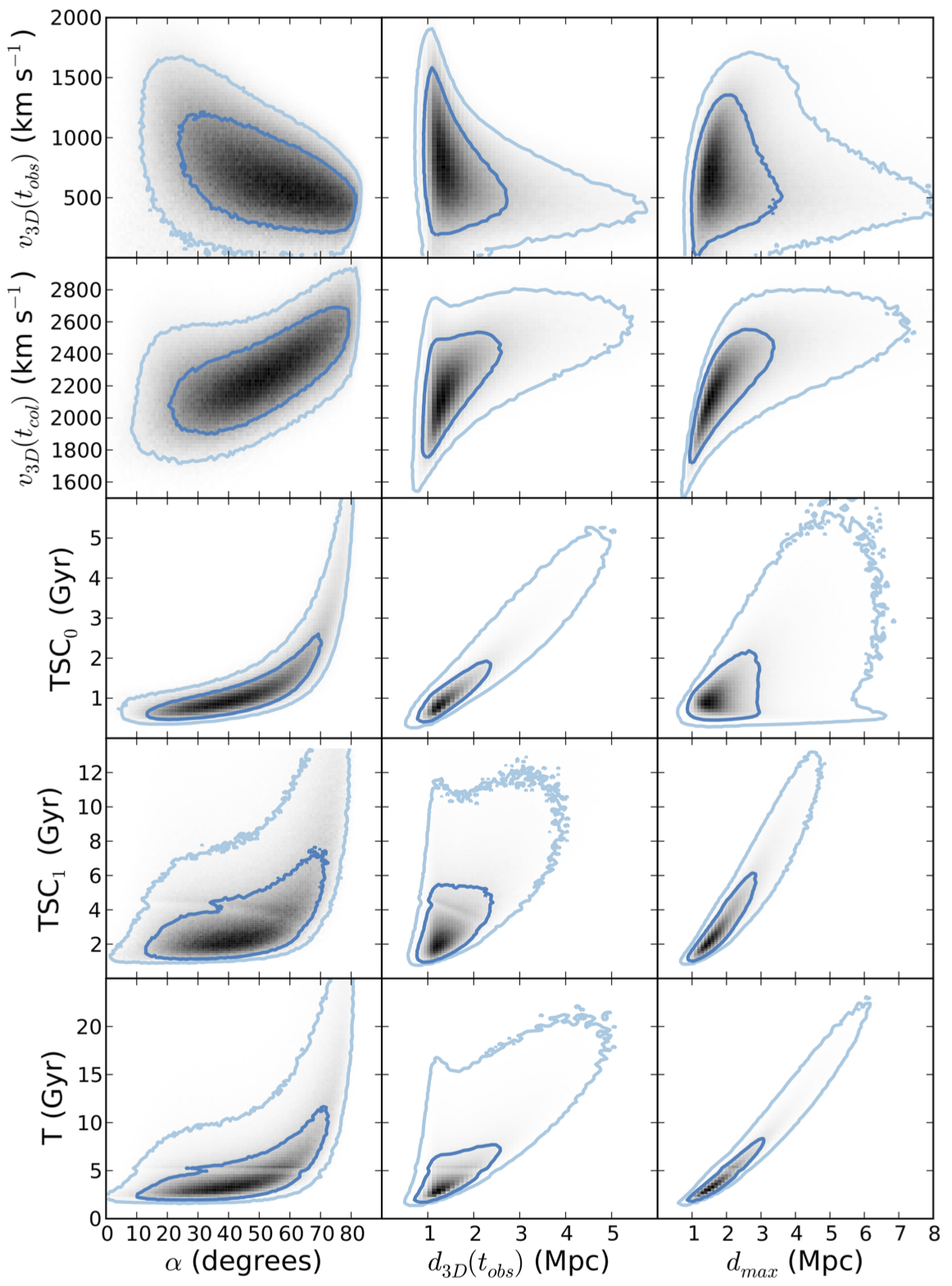}
\caption{Musket Ball Cluster marginalized \emph{Geometry vs.\,Velocity \& Time} parameters result plots.  Dark and light blue colors correspond to 68\% and 95\% confidence intervals, respectively.
\label{musket_geovt}}
\end{figure}

\begin{figure}
\epsscale{1}
\plotone{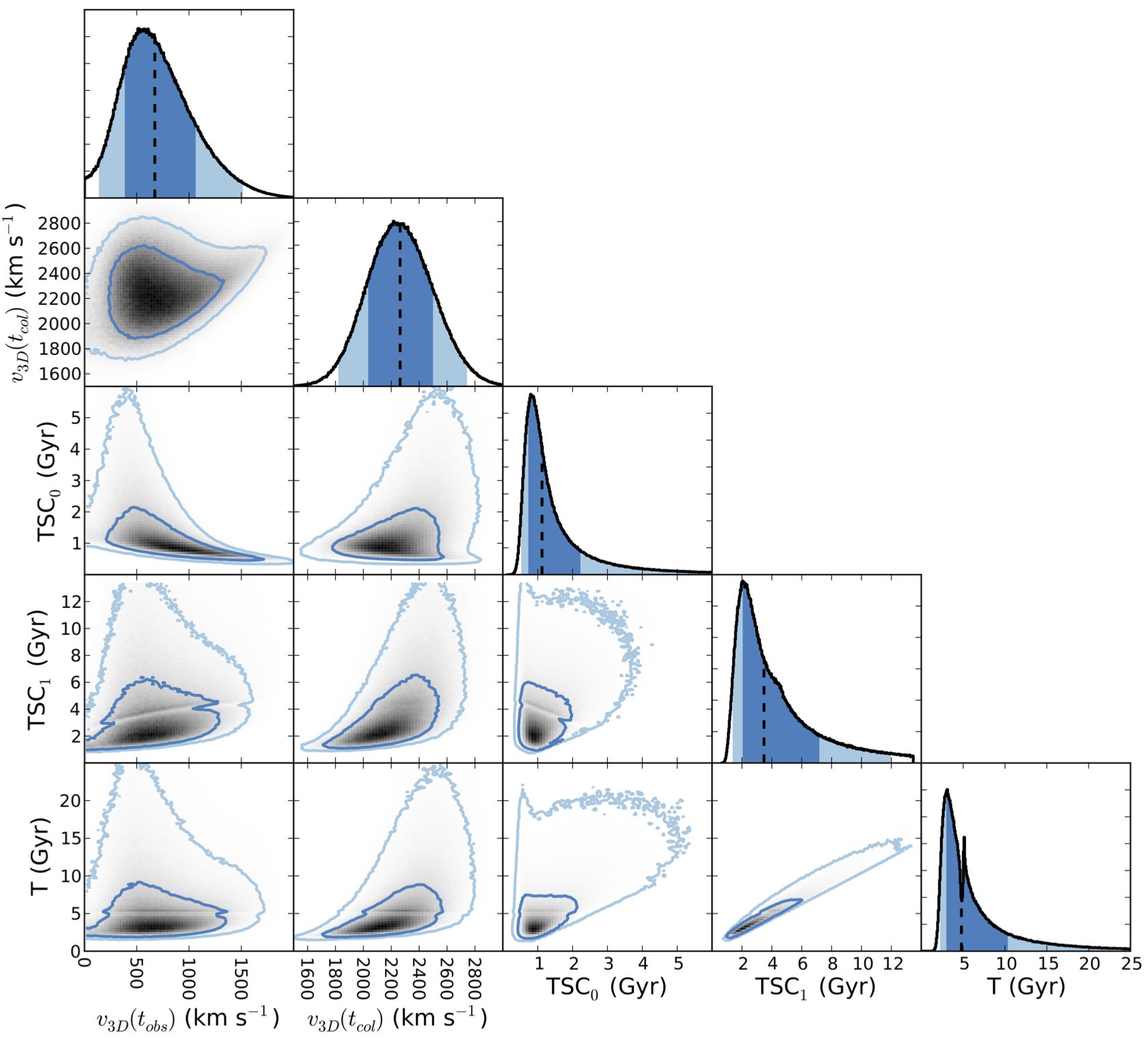}
\caption{Musket Ball Cluster marginalized \emph{Velocity \& Time vs.\,Velocity \& Time} parameters result plots.  Dark and light blue colors correspond to 68\% and 95\% confidence intervals, respectively.  The black dashed line is the biweight-statistic location \citep{Beers:1982dp}.
\label{musket_vtvt}}
\end{figure}
\clearpage								

\bibliographystyle{apj}
\bibliography{MCdyn}{}


\end{document}